\documentclass[prb,twocolumn,showpacs,preprintnumbers,amsmath,aps]{revtex4}
\usepackage{graphicx,hyperref}
\usepackage{amssymb}
\usepackage{bm}
\usepackage{subfigure}

\newcommand{\tr}{\mbox{Tr}}
\newcommand{\lrg}[1]{\left\langle #1 \right\rangle}
\newcommand{\atz}[1]{\left. #1 \right|_0}
\newcommand{\supsubdue}[2]{\begin{smallmatrix} #1 \\ #2 \end{smallmatrix}}
\newcommand{\supsubtre}[3]{\begin{smallmatrix} #1 \\ #2 \\ #3 \end{smallmatrix}}

\begin{document}

\title{On the fragility of the mean-field scenario of structural glasses for finite-dimensional disordered spin models}

\author{Chiara Cammarota} \email{chiara.cammarota@cea.fr}
\affiliation{IPhT, CEA/DSM-CNRS/URA 2306, CEA Saclay, F-91191 Gif-sur-Yvette Cedex, France}
\affiliation{LPTMC, CNRS-UMR 7600, Universit\'e Pierre et Marie Curie,
bo\^ite 121, 4 Pl. Jussieu, 75252 Paris c\'edex 05, France}

\author{Giulio Biroli} \email{giulio.biroli@cea.fr}
\affiliation{IPhT, CEA/DSM-CNRS/URA 2306, CEA Saclay, F-91191 Gif-sur-Yvette Cedex, France}

\author{Marco Tarzia} \email{tarzia@lptmc.jussieu.fr}
\affiliation{LPTMC, CNRS-UMR 7600, Universit\'e Pierre et Marie Curie,
bo\^ite 121, 4 Pl. Jussieu, 75252 Paris c\'edex 05, France}

\author{Gilles Tarjus} \email{tarjus@lptmc.jussieu.fr}
\affiliation{LPTMC, CNRS-UMR 7600, Universit\'e Pierre et Marie Curie,
bo\^ite 121, 4 Pl. Jussieu, 75252 Paris c\'edex 05, France}

\date{\today}

\begin{abstract}
At the mean-field level, on fully connected lattices, several disordered spin models have been shown to belong to 
the universality class of ``structural glasses'', with a ``random first-order transition'' (RFOT) characterized by 
a discontinuous jump of the order parameter and no latent heat. However, their behavior in finite dimensions  is 
often drastically different, displaying either no glassiness at all or a conventional spin-glass transition. We 
clarify the physical reasons for this phenomenon and stress the unusual fragility of the RFOT to \textit{short-range} 
fluctuations, associated \textit{e.g.} with the mere existence of a finite number of neighbors.  Accordingly, the 
solution of fully connected models is only predictive in very high dimension whereas, despite being also mean-field 
in character, the Bethe approximation provides valuable information on the behavior of finite-dimensional systems. 
We suggest that before embarking on a full-blown account of fluctuations on all scales through computer simulation or 
renormalization-group approach, models for structural glasses should first be tested for the effect of short-range 
fluctuations and we discuss ways to do it. Our results indicate that disordered spin models do not appear to pass 
the test and are therefore questionable models for investigating the glass transition in three dimensions. This also 
highlights how nontrivial is the first step of deriving an effective theory for the RFOT phenomenology from a rigorous 
integration over the short-range fluctuations.
\end{abstract}

\maketitle

\section{Introduction}
\label{introduction}

The random first-order transition (RFOT) theory\cite{KTW,wolynesbook} of the glass transition builds on a mean-field 
scenario, in which a complex free-energy landscape with an exponentially large number of metastable states emerges 
below a critical temperature $T_d$ and  the configurational entropy associated with these metastable states vanishes 
at a lower temperature $T_K$. At $T_K$  a RFOT, \textit{i.e.} a transition with a discontinuous order parameter yet 
no latent heat, to an ideal glass takes place. This scenario is realized in mean-field-like approximations of a 
variety of glass-forming systems (liquid models, lattice glasses, uniformly frustrated systems) as well as in mean-field, 
fully connected, spin models with quenched disorder \cite{wolynesbook,Cavagnareview} (\textit{e.g.}, $p$-spin model, Potts glass). 
The latter correspond to spin glasses without spin inversion symmetry, and their behavior differs from that of the fully 
connected Sherrington-Kirkpatrick Ising spin glass which is characterized by a 
continuous transition in place of the RFOT. They have been investigated in great detail and have provided most of 
the clues about the generic behavior of  ``mean-field structural glasses''.

To make progress toward a theory of the glass transition, one must however go beyond the mean-field description and 
include the effect of fluctuations\cite{CBTT} in finite-dimensional systems with finite-range interactions. The main 
assumption behind the RFOT theory of glass formation is that the mean-field scenario with a dynamic and a static 
critical temperature retains some validity when fluctuations are taken into account. The ergodicity breaking transition 
at $T_d$ is expected to be smeared and the metastable states no longer have an infinite lifetime because of entropically 
driven nucleation events. This underlies the picture of a glass-forming liquid as a ``mosaic state'' with its relaxation 
to equilibrium  dominated by thermally activated rare events involving ``entropic droplets''\cite{KTW}. Yet, the main 
ingredients associated with the RFOT are assumed to persist, even if renormalized by the effect of the 
fluctuations\cite{wolynesbook}. The issue can be understood in the much simpler setting of the 
liquid-gas transition of simple fluids. The mean-field van der Waals approach predicts a liquid-gas transition with a 
terminal critical point. It is known that this homogeneous mean-field picture needs to be modified, \textit{e.g.} 
through the classical nucleation theory and the renormalization group: concepts such as metastability and spinodal are 
no longer crisply defined, critical exponents as well as nonuniversal quantities are modified, yet the transition with the 
two, liquid and gas, free-energy states remains valid\dots  at least when the dimension $d$ is larger than $1$. In $d=1$ 
the mean-field treatment is plain wrong and predicts a transition that is not present. Fluctuations can therefore have a
more or less dramatic influence on the mean-field scenario: this is the key-point  concerning the relevance of the RFOT 
theory to glass-forming liquids.\\
One natural path to follow in order to investigate the effect of the fluctuations on the RFOT and the two-temperature 
picture is then to consider, both numerically and analytically, the various proposed models of structural glasses in 
finite dimensions. The disordered spin 
models are especially convenient as they are both well-defined at the mean-field level and much easier to investigate 
than more realistic models of structural glass-formers or effective Ginzburg-Landau theories in the replica formalism. 
In particular, they quite directly lend themselves to computer simulations and to real-space renormalization group (RG) 
treatments. The major obstacle on this seemingly straightforward route is that so far no traces of the RFOT scenario have 
been found in such studies on finite-dimensional, finite-range models, with either a complete absence of  transition to a 
glass phase \cite{brangian02} or a behavior more compatible with a continuous spin-glass transition\cite{pottsSG1} than a 
discontinuous, ``random first-order'', one.

In this work we clarify the reasons for the discrepancy between the solution of mean-field fully connected disordered spin 
models and their finite-dimensional behavior. All fully connected models that we have studied (Potts glasses, $M-p$ spin models 
and generalizations) provide predictions that turn out to be correct in very high dimensions only: because of their 
infinite connectivity, they indeed neglect local fluctuations and strongly enhance frustration compared to their 
finite-dimensional counterpart.\\
This stresses the importance of \textit{local, short-range} fluctuations on the existence of the RFOT. If one envisages 
integrating out the effect of the fluctuations in a renormalization group setting, it is often possible to proceed in a 
stepwise manner. First, one includes the short-range fluctuations up to a given scale and derives in this way an effective 
theory that describes the physics on longer distances (or lower energies). The next stage is then to solve the 
effective theory by accounting for the \textit{long-range} fluctuations, a usually highly nontrivial task. In the study of 
critical phenomena and phase transitions, the first step is in general bypassed and one relies instead on an effective 
Ginzburg-Landau model based on general symmetry arguments or on the mean-field solution of the microscopic model to 
which gradient terms are added in order to allow for spatial fluctuations. Generically, the type of mean-field theory 
considered does not matter. However, it \textit{does} matter in all of the 
disordered spin models that we have studied. As we shall show, the solution on fully connected lattices or in infinite dimension is not 
representative of the situation in finite-dimensional systems because it misses the effect of the (purely) short-range 
fluctuations associated with the mere existence of a limited number of neighbors. On the other hand, a better account of 
the latter is provided by the Bethe approximation, which, despite being mean-field in character, properly describes the 
local effect of a finite connectivity. The case of the Potts glass, which we shall discuss in detail, is paradigmatic. In 
their computer simulation study, Brangian et al \cite{brangian02} found no transition and no glassy behavior for a $10$-state 
Potts glass model in three dimensions, contrary to what was obtained in infinite dimensions. We show here that this is 
actually expected on physical grounds and confirm this through the solution of the model within the Bethe approximation.

The outcome of our study is two-fold: (1) stressing the importance to test the influence of short-range 
fluctuations on models of structural glasses through the Bethe approximation or similar approaches and (2) showing 
that disordered spin models mostly appear to fail the test and are therefore questionable starting points 
for investigating the glass transition in three dimension.

\section{From infinite to finite dimension: difficulties and puzzles}
\label{sec2}

On the way to finding a RFOT in finite-dimensional disordered spin models, several difficulties have been identified:

\textit{(1) The RFOT can give way to a continuous transition to a spin-glass, with a quite different phenomenology.}
This phenomenon was observed for several models when studied by numerical simulations in finite dimensions, see e.g. 
Ref. [\onlinecite{pottsSG1}].  The lack of evidence for the entropy crisis predicted by the RFOT theory in 
finite-dimensional disordered spin models has been rationalized by invoking the fact that, already at the
mean-field level, the relative temperature interval between $T_d$ and $T_K$ is very small for ``reasonable values'' 
of the number $p$ of irreducibly interacting spins in the $p$-spin glass model or of the number $q$ of states in the 
Potts glass model (say, $p,q \lesssim10$): under such conditions, the phenomenology should then look more like that 
of a standard Ising spin glass than that coming with an RFOT. Eastwood and Wolynes\cite{eastwood02} have related this 
to the small value of the reduced surface tension between glassy states in such models and argued that the $p$'s and $q$'s 
that would be necessary to mimic structural glasses to be $p\simeq 20$ and $q\simeq 1000$.

\textit{(2) The RFOT to an ideal glass can be superseded by a conventional (usually ferromagnetic) ordering transition.}
This phenomenon takes place for instance in the disordered Potts models where a tendency to a simple 
nonglassy ferromagnetic order was observed. Actually, this is also what happens in more realistic 
models of supercooled liquids in which there is always a transition to the crystal. In the latter case, 
however, since the transition is first-order one can supercool the liquid and study the metastable glassy phase. 
This in turn is impossible if the glass transition is superseded by a second-order phase transition, as the above 
mentioned ferromagnetic one.   

Either as a result of the above difficulties or in trying to avoid them, it is often found that no remnants of the RFOT are present in finite 
dimensions. Actually, it may also be that no transitions at all are observed if one has succeeded in getting around points (1) and (2).
An example is provided by the model studied by Brangian \textit{et al.}\cite{brangian02}. These authors considered a 
disordered Potts glass with a distribution of couplings that is displaced toward the antiferromagnetic ones and that, 
as a result, has only a small fraction of ferromagnetic couplings. This is a general procedure to avoid  ferromagnetic 
ordering in models without spin inversion symmetry. (An alternative suggestion is to study an all-ferromagnetic or 
all-antiferromagnetic Potts model with random permutations of the $q$ states; the latter enforce a statistical gauge symmetry 
that prevents ferromagnetic or antiferromagnetic ordering\cite{marinari99,randompermutationnumerics,krzakala07}.) 
In practice, Brangian et al. focused on a $10$-state Potts glass model with a bimodal distribution of the couplings 
having only a fraction $x\simeq 15\%$ of ferromagnetic (positive) ones. Their main result, which is often cited as a 
major problem for the RFOT theory, is that the three-dimensional model is not 
glassy at all, despite the fact that the mean-field solution in infinite dimension predicts a strong RFOT transition 
for such a large number of states. This, and further analytical and numerical work, led for instance 
Moore et al\cite{moore} to put forward the drastic proposal that the RFOT never survives in three dimensions, 
its phenomenology being instead replaced by that of a conventional spin glass in a magnetic field (which has no 
transition in three dimensions according to the droplet theory\cite{droplet_bray,droplet_fisher}). \\
In the following, we investigate the robustness of the RFOT in spin models with quenched disorder that display such 
a transition in their mean-field fully connected limit. We  clarify how the above listed difficulties arise when lowering 
the dimension from infinity down to three. We restrict ourselves to systems with short-range interactions involving 
only pairs of nearest-neighbor sites on a lattice. This choice is motivated 
by the fact that such models are amenable to real space RG analyses and that they are easier to study numerically 
in comparison to systems where multi-site interactions are present, such as the $p$-spin model. 
We first focus on Potts glasses and show that by trying to avoid problems (1) and (2) one actually imposes 
contradictory requirements in finite dimensions: when the distribution of couplings is mostly antiferromagnetic, 
spontaneous ferromagnetic ordering is indeed thwarted, but considering large values  of the number of states $q$ 
then suppresses frustration and leads to no glass transition at all. We then analyze two alternative types of models: 
the $M-p$ spin models introduced in Ref. [\onlinecite{mpmodel}] and a new class introduced by us. We discuss the 
physical reasons that make the predictions based on the solution of the infinite-dimensional model inadequate in 
low dimensions, pointing out, as already mentioned, the importance of short-range fluctuations. 

\section{Disordered Potts models}
\label{potts}

This section is devoted to a systematic study of disordered Potts models on lattices with finite connectivity.
Our aim is to understand the physical reasons that make the discontinuous glass transition (RFOT) 
so elusive for these systems. We shall show that the two conflicting requirements discussed above
actually put severe constraints on the dimensionality and the connectivity of the possible lattices.   

\subsection{Models}

We define in the following the two Potts models that we shall focus on:

{\it (1) Potts glass with $q$ states}. Its Hamiltonian is given by
\begin{equation}
H=-q \sum\limits_{\langle i,j \rangle} \, \, J_{ij}\, \delta_{\sigma_i \sigma_j}  \quad ,
\label{eq1}
\end{equation}
where the coupling constants $J_{ij}$ are quenched random variables,
and the $\sigma_i$'s are $N$ Potts variables that can take $q$ different values, $\{1,2,É,q\}$. 
The symbol $\langle i,j \rangle$ means that each pair of nearest-neighbor sites $i,j$ on the lattice 
is included in the sum only once and 
$\delta_{\sigma \sigma'}$ is the Kronecker symbol. 
We consider bimodal and Gaussian distributions of the coupling constants.
The mean-field solution of this model was worked out in Ref. [\onlinecite{sompolinsky}] and depends only
on the  mean and the variance of the coupling constant distribution, denoted $J_0$ and $(\Delta J)^2$ respectively.
By scaling $J_0$ and $(\Delta J)^2$ with the spatial dimension $d$ as
\begin{equation}\label{couplings1}
J_0=\frac{\hat J_0}{d}\qquad (\Delta J)^2=\frac{1}{2d},
\end{equation}
where for convenience we use the rescaled $\Delta J$ as the unit of energy and temperature, one obtains a well 
defined mean-field model in the limit $d\rightarrow \infty$. (The same is true in the limit $N\rightarrow \infty$ 
when considering a completely connected lattice and replacing $d$ with the number of sites $N$.)

A detailed presentation of this model is postponed to the following section. 
Here we just recall the main result obtained within mean-field theory. Note that we only focus
on the case $q>4$. This corresponds to models displaying a dynamical ergodicity-breaking transition at a temperature
$T_d$ and a RFOT at a temperature $T_K$. The former is in the class of the singularity found in the mode-coupling theory 
of liquids and the latter is akin to an entropy-vanishing, Kauzmann-like, discontinuous glass transition. Lower values 
of $q$ lead to a continuous spin-glass transition\cite{pottsSG1}.  As discussed before, it is important to take 
$\hat J_0$ negative enough in order to avoid ferromagnetic ordering. In practice, this means 
$\hat J_0<-\frac{p-4}{2}\frac{1}{2T}$.
The dynamical and static transition temperatures depend on $q$ as $T_d\sim \sqrt{\frac{q}{2\log q}}$ and 
$T_K\sim \frac 1 2 \sqrt{\frac{q}{\log q}}$ for $q\gg 1$ \cite{caltagirone}. The values of the jump 
in the (overlap or spin-glass) order parameter at $T_d$ and $T_K$ both tend to one in the large $q$ limit, 
indicating that the larger the number of states the more discontinuous the transition at $T_K$.

{\it (2) Random-permutation Potts magnet}. Its Hamiltonian 
reads  
\begin{equation}
\label{eq_hamiltonian}
\begin{aligned}
H= J \sum_{<i,j>}\delta_{\sigma_i, \pi_{ij}(\sigma_j)},
\end{aligned}
\end{equation}
where $J>0$ (antiferromagnet) or $J<0$ (ferromagnet) and $\pi_{ij}$ is a random permutation of the $q$ colors 
that is attached to the edge between $i$ and $j$\cite{footnote1}.  This model has a ``gauge invariance'' 
that prevents antiferromagnetic or ferromagnetic 
ordering\cite{marinari99,randompermutationnumerics,krzakala07}: indeed, if one permutes the states of the spin at a 
given site $i$, one can always find random permutations associated with all  edges emanating from $i$ such that the 
energy does not change; after averaging over the quenched disorder (\textit{i.e.} random permutations), the (staggered) 
magnetization is then zero.
This model has been studied by numerical simulation\cite{randompermutationnumerics} in its ferromagnetic version and 
by the cavity method\cite{krzakala07} in its antiferromagnetic version. In the latter case, the $T=0$ limit describes 
a form of ``coloring problem'', the states being then interpreted as colors.

\subsection{Frustration and lack thereof for $q\gg1$}

In the search for finite-dimensional models displaying glassy phenomenology, the disordered Potts models have played
a central role. They were actually at the root of the RFOT theory of the glass transition in the pioneering papers 
by Kirkpatrick, Thirumalai and Wolynes \cite{KTW}. For this reason and also because the mean-field studies cited 
above predicted a strong discontinuous glass transition for a large number of states (or colors) they seemed to 
be the most promising candidates to find a RFOT in three dimensions. It thus came as a surprise that simulation 
results\cite{brangian02} instead showed a complete absence of  glassy phase for a $q=10$ disordered Potts model.

Our claim, which we substantiate in the following, is that the original intuition based on the infinite-dimensional limit
(or the study of fully connected lattices) is misguided. A first piece of explanation comes by considering the degree 
of frustration, which is known to be central for the existence of glassy phases. This is most clearly understood in 
the case of the disordered antiferromagnetic Potts model, which corresponds to a coloring problem at zero temperature. 
For any {\it finite-connectivity} lattice, such a model becomes unfrustrated for $q$ large enough: in physical terms, 
if the number of states, \textit{i.e.} colors,  is 
too large for a given coordination number then it becomes easy to arrange them in a way that two neighboring sites 
do not have the same color, hence lifting frustration. Therefore, one does not expect any glassy phase for large $q$ 
on a finite-connectivity, and in particular a finite-dimensional, lattice. 

The recent study of the antiferromagnetic Potts model with random color permutations on (both regular and Erd\"os-R\'enyi) 
random graphs of finite coordination number $c$\cite{krzakala07} supports our claim. It was shown that the model for 
$q\geq 4$ may have a phenomenology similar to that of other mean-field structural glasses, with a RFOT accompanied by 
a dynamical transition at $T_d$. However, in agreement with our previous discussion, the RFOT disappears at fixed 
coordination number $c$ when the number of colors $q$ is sufficiently large: for instance, the transition disappears 
for $c \leq 8$ for $q=4$ and for $c\leq 13$ there is no phase transition already with $q=5$. Asymptotically, the 
transition is absent when $2q\log(q) - \log(q)-2\log(2)>c$ (see footnote in Ref. [\onlinecite{footnote2}]). For a given lattice coordination number, 
frustration, hence glassy phenomenology, is therefore lost when $q$ is too large, typically larger than $(c/2)/\log(c/2)$. 

The same trend takes place in Euclidean lattices. We indeed prove in appendix \ref{dobrushin} that the $q$-state 
antiferromagnetic Potts model with random color permutations has no phase transition on Euclidean lattices when $q>2c$. 
This generalizes a proof by Salas and Sokal \cite{salas97} that applies to the standard random antiferromagnet Potts model. 
In particular we show that there is a unique infinite-volume Gibbs measure with exponential decay of the correlations at 
all temperatures and for any realization of the random permutations. Quite generally then, frustration vanishes for 
a large number of colors or states, as intuitively expected. 

\subsection{Bethe approximation of the ten-state three-dimensional Potts glass model}

In the previous section we have shown that frustration vanishes when the number of states $q$ becomes sufficiently 
large at fixed coordination number in random antiferromagnetic Potts models on both Euclidean lattices and random graphs. 
As a result, any putative glassy phase transition is wiped out. We now study the effect of adding a 
fraction of ferromagnetic couplings. This should introduce frustration and possibly lead to RFOT phenomenology. 
We shall see, however, that this is not the case.

\begin{figure}
\includegraphics[angle=-90,width=.5\textwidth]{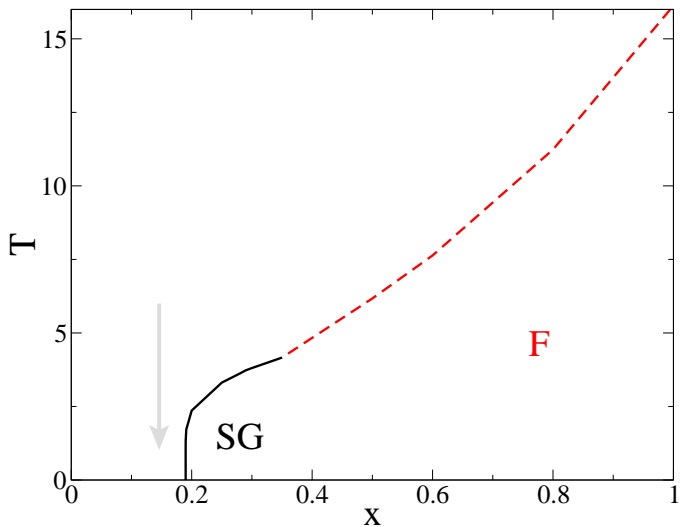}
\caption{Phase diagram of the ten-state three-dimensional Potts glass with bimodal disorder obtained within the 
Bethe approximation. The value of $x$ shown by the arrow corresponds to that studied by Brangian et al.\cite{brangian02}. We 
have only drawn the transition lines with the paramagnetic state. There is also a transition line between the ferromagnetic 
and the spin-glass state which we do not report.}
\label{binder}
\end{figure}

In the following we focus on the model considered by Brangian \textit{et al.}\cite{brangian02}, which we have 
investigated through the Bethe approximation: in practice, we have then studied the $q=10$ Potts glass on a random graph 
with the same coordination number as the cubic lattice, $c=6$. We have used the cavity method that allows for an 
analytic solution. In this model, the distribution of the couplings is bimodal, 
\begin{equation}
\label{couplings}
P(J_{ij})=x \delta(J_{ij}-J) + (1-x) \delta(J_{ij}+J),
\end{equation}
where $J=\sqrt 2$ and the fraction of ferromagnetic couplings $x$ varies from 0 to 1. The value considered by 
Brangian \textit{et al.} is $x=(2-\sqrt 2)/4\simeq 0.146$\cite{brangian02}. The phase diagram that we have obtained 
within the Bethe approximation is reported in Fig. \ref{binder}. There is no discontinuous glass transition (RFOT) 
for any positive fraction $x$ of ferromagnetic couplings. Instead, for relatively small values of $x$ we find a continuous 
spin-glass transition and, above some threshold, this spin-glass transition is superseded by a ferromagnetic one.   
The model is not glassy at all for $x=(2-\sqrt 2)/4\simeq 0.146$, at any temperature. Although there is no general 
proof, one expects that frustration on Euclidean lattices is comparable or less than that found on random graphs. 
Thus, the absence of a RFOT on a random graph for $x=(2-\sqrt 2)/4\simeq 0.146$ is fully compatible with the numerical 
results of Brangian \textit{et al.}.

From this analysis, we therefore conclude that adding a small fraction of positive couplings is not sufficient 
to trigger the appearance of a discontinuous glass transition when the connectivity is finite and the number of 
states large. In addition, this shows that the absence of RFOT in the three-dimensional model studied by 
 Brangian \textit{et al.} is not primarily due to some long-distance, possibly non-perturbative, fluctuations but 
 is just a consequence of the local property
 of the lattice, namely the fact that the connectivity is too small compared to the number of colors. This effect 
 is correctly captured by the Bethe approximation. It is instead completely missed by the mean-field analysis based on a 
 fully connected lattice, which therefore appears to be quite misleading to predict the behavior of three-dimensional systems.
  
An additional question that could provide a valuable hint for future studies concerns the smallest value of the spatial dimension $d$ 
(and of the associated number of states $q$) such that Potts disordered models show 
a discontinuous glass transition (RFOT) within the Bethe approximation. We consider as a prototypical example the case of the 
Potts model with the distribution of the couplings given by Eq.~(\ref{couplings}) with $q=5$ states and placed on a 
regular random graph of coordination number $c=18$ 
(which would mimic a hypercubic lattice in $9$ dimensions).

\begin{figure}
\includegraphics[angle=-90,width=.5\textwidth]{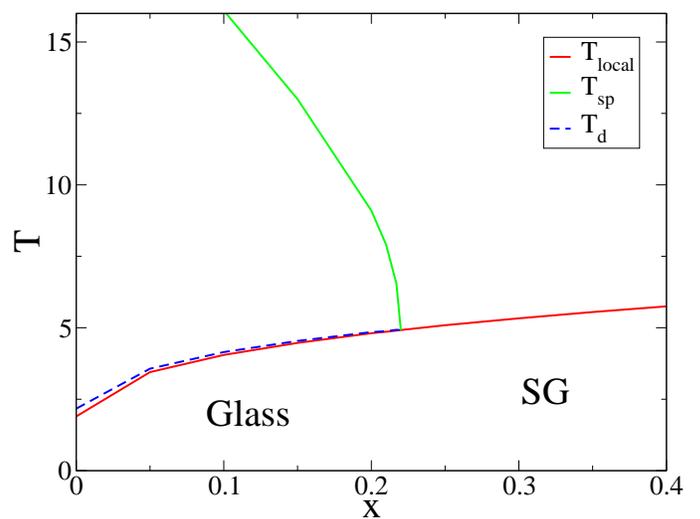}
\caption{Phase diagram of the five-state Potts glass with bimodal disorder on 
a regular random graph of coordination number $c=18$ as a function of $x$ and $T$, 
showing $T_{sp}$ (green line), where the paramagnetic phase becomes unstable towards 
the antiferromagnetic order, $T_{local}$ (red line), where the RS solution becomes
unstable towards RSB, and the dynamical temperature $T_d$ (blue dotted line). We do not know if the 
virtual coincidence of the points at which $T_{loc}$ merges with $T_d$ on the one hand and with 
$T_{sp}$ on the other is a result of the numerical uncertainty or has a deeper significance.} 
\label{fig:q5}
\end{figure}

The phase diagram of the model, which we have obtained through 
the cavity method, is shown in Fig.~\ref{fig:q5} as a function of $x$ and $T$.
For a small enough $x$ ($x\le 0.22$), where most of the couplings are 
antiferromagnetic, there is enough frustration to produce a RFOT,
as previously shown for $x=0$\cite{krzakala07}. This is demonstrated by the fact that the dynamical 
critical temperature, $T_d$, is higher than the temperature $T_{local}$ at which the 
replica-symmetric (RS) solution becomes unstable towards replica-symmetry-breaking (RSB).
However, the glass transition is superseded by an antiferromagnetic transition, which takes
place at a much higher temperature. We have computed the temperature $T_{sp}$ at which the paramagnetic 
solution becomes unstable towards antiferromagnetic order\cite{footnote3}: see Fig.~\ref{fig:q5}. This transition 
will likely be very difficult to avoid in any numerical simulation on Euclidean lattices. As the value of $x$ is increased, 
$T_{sp}$ decreases, but, at the same time, the distance between $T_{local}$ and
$T_d$ reduces and the glass transition becomes less and less discontinuous.
Finally, when $x>0.22$, where there is no instability towards the antiferromagnetic order,  
the glass transition becomes continuous and a conventional spin-glass phase is found.\\
This example confirms that even in a dimension as high as $d=9$, avoiding the two problems listed in Sec. \ref{sec2} imposes 
conflicting constraints that are very hard to fulfill. Some improvement could come from considering random permutations in order 
to avoid the appearance of an antiferromagnetic phase. Even in this case, one would have to consider for $q=4$ at least $c\geq 12$, 
\textit{i.e.} $d\geq 6$ for a hypercubic lattice.
\section{The approach to the infinite-dimensional or fully connected limit}
\label{limit}

In this section we study the $1/d$ expansion for the $q$-state Potts glass on Euclidean hypercubic lattices of 
coordination number $c=2d$. The aim of this analysis is to better understand the regime of validity of the 
mean-field results which, as found in the previous section, appears to be rather limited.

We have derived the $1/d$ expansion of the replicated Gibbs  free energy by following the method developed by 
Georges, M\'ezard and Yedidia\cite{yedidia1,yedidia2}. We have focused on a Gaussian distribution of couplings.
In order to perform the $1/d$ expansion it is useful to introduce the simplex representation of the model, as follows:
\begin{equation}\label{Hsimplex}
H=-\sum_{<i,j>}{J}_{ij} \sum_{a=1}^{q-1} S_{i,a}S_{j,a} \ .
\end{equation}
In this representation $q$ is the number of colors and the degrees of freedom $S_{i,a}$ are vectors 
pointing toward the $q$ vertices of a tetrahedron in a $(q-1)$-dimensional space \cite{cwilich}. 
The order parameter within the replica treatment reads:
\begin{equation} 
Q_i^{\alpha\beta} =\lrg{S^{\alpha}_{i,a} S^{\beta}_{i,a}}=\lrg{\delta_{\sigma^{\alpha}_{i} \sigma^{\beta}_{i}}}
\label{q}
\end{equation}
where $\alpha$ and $\beta$ are replica indices. We refer to Ref. [\onlinecite{cwilich}] for more details on the replica theory 
for Potts glasses. When using the simplex representation, the Hamiltonian of the Potts model is very similar to that 
of the Edwards-Anderson (Ising) model. The $1/d$ expansion can then be performed through the high-temperature expansion 
of the replicated Gibbs free energy\cite{yedidia1}. In the case of Potts variables the computation is more cumbersome. 
For this reason we only consider the first order in $1/d$. The detailed computation is presented in appendix B and 
in the following we only report the main results. 

The Gibbs free-energy $A$ can be obtained as an expansion in $Q^{\alpha \beta}$. We focus
on the first three terms only, since this is enough to discuss the existence and the properties of the glass transition: 
\begin{equation}\label{effectiveF}
\begin{split}
&-\beta A\simeq \\& N\Biggl[\Biggr.
\sum_{\alpha\neq\beta}-\frac{t}{4}(Q^{\alpha\beta})^2+
\frac{w_1}{6}\tr(Q^3)+\sum_{\alpha\neq\beta}\frac{w_2}{6}(Q^{\alpha\beta})^3+...\Biggl.\Biggr] \ .
\end{split}
\end{equation}
The coefficients of the expansion computed at the first order in $1/d$ read
\begin{equation}
t=1-\beta^2-\frac{\beta^4}{4d}(q^2-10q+10)-
2(q-2)\beta^3\frac{\tilde{J}_0}{d}-2\frac{(\beta\tilde{J}_0)^2}{d}\nonumber
\end{equation}
\begin{equation}\label{w1w2Potts}
w_1=1-\frac{3\beta^4}{2d}(q-1)\qquad,\qquad w_2=\frac{q-2}{2}
\end{equation}
where 
\begin{equation}
\beta \tilde{J}_0=\left(\beta \hat J_0+\frac{\beta^2(q-2)}{2}\right) \ .
\end{equation}
As in the mean-field solution ($d\rightarrow \infty$), we consider a one-step replica symmetry broken (1-RSB) form 
for $Q^{\alpha \beta}$: the replicas are 
grouped in $n/m$ groups of $m$ elements and $Q^{\alpha \beta}$ is equal to $Q$ for all pairs belonging to the same group
and to zero otherwise. In order to study the glass transition one has to focus on the $m\rightarrow 1$ limit 
of $(\beta A)/(m-1)$ which, within the 1-RSB ansatz, reads 
\[
\lim_{m\rightarrow 1}\frac{\beta A}{m-1}=\frac t 4 Q^2-\frac{w_2-w_1}{6} Q^3+...\]
Within this formalism a discontinuous glass transition (\textit{i.e.} a RFOT) is signaled by the sudden change of the global 
minimum 
of $(\beta A)/(m-1)$ from $Q=0$ to $Q=Q_{EA}>0$. 
A negative sign of the coefficient of the cubic term, {\it i.e.} $w_2>w_1$, 
then insures that the transition, if it exists, must be discontinuous. In addition, if the coefficient of the quadratic 
term, $t/4$, 
becomes negative below a temperature $T_{local}$, then a discontinuous transition surely takes place for $T\ge T_{local}$. 
In the strictly infinite-dimensional limit, one finds $w_2>w_1$ for $q>4$ and $T_{local}=1$, in agreement with the 
exact solution of the model\cite{sompolinsky}. (One also has to consider $\tilde J_0$ equal to zero or negative in 
order to avoid ferromagnetic ordering.) The magnitude of the coefficient of the cubic term increases (\textit{i.e.} 
the term becomes more negative) with the  number of states. This fact and the exact mean-field solution showing that 
$Q_{EA}\rightarrow 1$ for $q\rightarrow \infty$ led to the belief that the larger the number of states (or colors) 
the more strongly discontinuous the transition in finite dimension. 
The $1/d$ corrections also seem to support this conclusion. Indeed, at $T\simeq T_{local}$, when $d\gg q^2$, which 
corresponds to the perturbative regime of the $1/d$ expansion, the corrections make $w_2-w_1$ larger and $T_{local}$ 
higher when $q$ increases (if $\tilde J_0\le 0$). The latter condition implies that the scaled mean of the couplings, 
$\hat J_0$, is negative enough, \textit{i.e.} that  the model is antiferromagnetic enough. This is however quite surprising 
since for a $\hat J_0$ that is very negative  
the corresponding finite-dimensional model has almost all of the couplings antiferromagnetic. As we have shown 
in the previous section, increasing the number of Potts states leads to a decrease of frustration for such a system, 
and eventually a disappearance of RFOT, in finite dimensions. The $1/d$ expansion appears instead to predict just the opposite!

The disagreement between the two approaches (the study on random and Euclidean lattices for a large number of states and 
the $1/d$ expansion) can be traced back to the fact that in the former case one considers the limit of large number of 
states at fixed dimension whereas in the latter the number of dimension tends to infinity at fixed, although large, 
number of states.  These two different ways of taking the limits of large dimension and of large number of states do 
not seem to commute. For instance, in the limit of large dimensions, the couplings are such that the average value is 
typically much less than the standard deviation: $\hat J_0/d\ll \Delta J=\sqrt{1/d}$ if $d$ is large enough, no matter 
what is the value of $\hat J_0$. In consequence, the starting point of the $1/d$ expansion is such that almost half of 
the couplings are positive whereas the other half are negative, a situation quite different from the 
finite-dimensional one we would like to describe, \textit{i.e.}  a Potts glass in which almost all of the couplings are 
antiferromagnetic.  

In conclusion, the mean-field infinite-dimensional theory appears to be rather singular. It may still be representative of 
the physics of finite-dimensional Potts glasses, but only for a very large number of dimensions, and certainly not for the 
three-dimensional case. (As discussed above, we expect that $d$ has to be larger than the square of the number of states in 
order  to be in the perturbative regime related to mean-field theory.)
As a consequence, the free energy in Eq. (\ref{effectiveF}) with the coefficients given by Eq. (\ref{w1w2Potts}) does not 
provide a correct effective model for the three-dimensional system even when the first $1/d$ corrections are included. 
  
\section{Alternative Models}

In the following we focus on two classes of disordered models, one introduced in Ref. [\onlinecite{parisipicco}] and one 
tailored by us, which might provide a better alternative to Potts glasses.  Unfortunately, as we shall show, the 
drawbacks of the disordered Potts models discussed above apply, at least partially, to these alternative models too. 

\subsection{The $M$-$p$-spin model}

The $M$-$p$-spin models were first introduced in Ref. [\onlinecite{parisipicco}] and later generalized and studied in Refs. 
[\onlinecite{romani,moore}]. $M$ Ising spins $S^{(\alpha)}_i$, $\alpha=1,2,\cdots,M$ are present on each site $i$
of a hypercubic lattice. In this work we consider the $p=3$ case, whose Hamiltonian is given in terms of
products of $3$ spins chosen from the spins in a pair of nearest-neighbor sites: 
\begin{eqnarray}
H&=&-\sum_{\langle ij\rangle}\sum^M_{\alpha<\beta}\sum^M_{\gamma}\Big( 
J^{(\alpha\beta)\gamma}_{ij} S^{(\alpha)}_i S^{(\beta)}_i 
S^{(\gamma)}_j \nonumber\\
&&\quad\quad\quad\quad\quad\quad +
J^{\gamma(\alpha\beta)}_{ij} S^{(\gamma)}_i S^{(\alpha)}_j 
S^{(\beta)}_j\Big)\ .
\end{eqnarray}
The couplings are independently distributed quenched Gaussian variables with zero mean and variance $\Delta J^2$. Note that 
the interactions involve three spins but only two sites.

The expansion of the Gibbs free energy in terms of $Q^{\alpha \beta}$, analogous to the one described above, 
was worked out in Ref. [\onlinecite{romani}] for general values of $M$ and $p$ and $d=\infty$. The unique value of $M$ 
such that $w_2>w_1$, {\it i.e.} leading to a discontinuous glass transition (RFOT), is $M=3$ for $p=3$. This model
was later studied by Migdal-Kadanoff real-space RG in three dimensions by Yeo and Moore \cite{moore}. 
They concluded that no glass transition  is present in finite dimension and that the behavior of the system resembles 
more that of a model displaying an avoided continuous spin-glass transition. There is still no numerical simulation testing 
this prediction (work is in progress \cite{romanisimulation}).

\begin{figure}
\includegraphics[angle=-90,width=.5\textwidth]{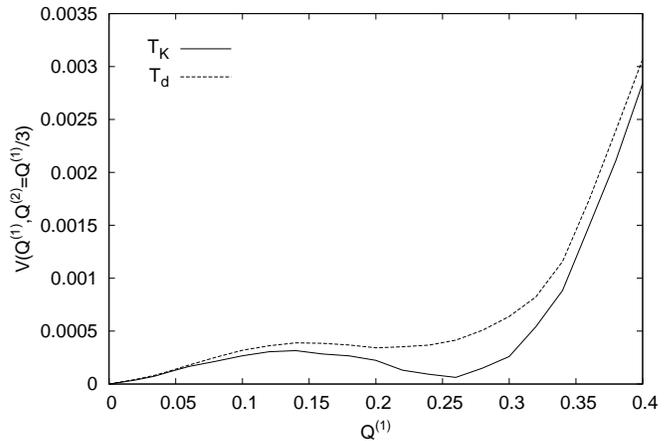}
\caption{Plot of $V=(\beta A)/(m-1)$ as a function of the overlap for the $M-p$ disordered model with 
$M=3$ and $p=3$ at $T_d$ (dashed line) and at $T_K$ (continuous line). As discussed in the main text and in Appendix C, the 
exact solution of the model lead to two kinds of overlaps (or order parameters) $Q^{(1)}$ and $Q^{(2)}$. We plot the Gibbs 
free energy as a function of $Q^{(1)}$ along the line 
$Q^{(2)}=Q^{(1)}/3$; a similar behavior is obtained for other choices.}
\label{figromani}
\end{figure}

In order to understand the disagreement between the two approaches (infinite-dimensional mean-field and real-space RG in $d=3$), 
we have solved the model on a fully connected lattice with the same choice of scaling for the variance
as in Ref. [\onlinecite{romani}]: $\Delta J^2=\frac{1}{9N}$ (this is equivalent to take $d=\infty$ and a variance of the 
coupling scaling as $1/d$). More details are given in appendix C. We have found that the dynamical MCT-like transition 
takes place at $T_d\simeq 0.5970$ and that the static RFOT takes place at $T_K\simeq0.5963$. The plot of the Gibbs 
free energy, $(\beta A)/(m-1)$, within the 1-RSB ansatz is shown in Fig. \ref{figromani} for $T=T_d$ and $T=T_K$. 
The most striking fact is the very small value of the free energy for the whole range of overlap. Actually, in the 
exact solution of the models two different overlaps emerge: 
$Q^{(1)}=\sum_i\langle S^{\alpha}_i\rangle^2/N$ and $Q^{(2)}=\sum_i\langle S^{\alpha}_iS^{\beta}_i\rangle^2/N$. We have 
plotted the free energy along the line 
$Q^{(2)}=Q^{(1)}/3$ but a similar behavior is obtained for other choices. 
As we discuss below, the smallness of the free-energy is possibly the cause
of the lack of robustness of the mean-field results as the finite-dimensional fluctuations not taken into account within 
mean-field theory are then overwhelming compared to the structure of $(\beta A)/(m-1)$ found in $d=\infty$.

We have not tried to develop a $1/d$ expansion for this model because in finite dimension, due to the lack of 
inversion symmetry, $S_i\rightarrow -S_i$, a nonzero value of the overlap $Q_0$ (in the 1-RSB scheme) emerges 
even at high temperature: an amorphous, but trivial, disordered magnetization profile is induced by the quenched disorder. 
As a consequence, the $1/d$ expansion is more involved (one should make an expansion in $Q-Q_0$).

\subsection{The F-model}

We now consider another class of models. As the previous one it is characterized by spins with nearest-neighbor 
interactions.The partition function of the model 
is defined as
\[
Z=\sum_{\{S_i\}}\prod_i \mu(S_i^1,S_i^2,S_i^3)\exp(-\beta H)
\]
where the spins can take the values $-1,0,+1$ and $\mu$ gives different weights to the $3^3$ states per site; $\mu$ 
is a kind of generalized fugacity. 
We focus on the case where the only allowed states are $\{1,1,1\}$, $\{1,-1,-1\}$, $\{-1,-1,1\}$, $\{-1,1,-1\}$ and 
$\{0,0,0\}$,
{\it i.e} $\mu=0$ for all other states. Moreover, we introduce $P$ as the ratio between $\mu(1,1,1)$ and $\mu(0,0,0)$ 
and we assume
that the first four allowed states are all characterized by the same value of $\mu$. (We have considered other variants 
which have all led to
similar results; thus, we just present the simplest one.)

The Hamiltonian of the model reads
\[
H=-\sum_{\langle i,j \rangle}\sum_a J_{ij}^a S_i^{a} S_j^a
\]
where the $J_{ij}^a$'s are independently distributed Gaussian random variables with zero mean and variance $\Delta J^2$.
A welcome characteristic of this model is that it has a kind of spin-inversion symmetry, if one flips, say, all 
$S_i^1,S_i^2$ the probability measure is invariant (likewise for $S_i^1,S_i^3$ and $S_i^2,S_i^3$). In consequence
the average magnetization is zero at high temperature, {\it i.e.} $Q_0$ is equal to zero.

\begin{figure}
\includegraphics[angle=-90,width=.5\textwidth]{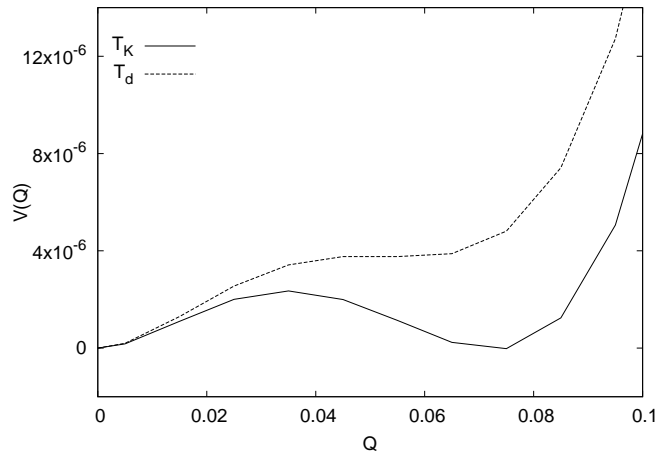}
\caption{Plot of $V=(\beta A)/(m-1)$ as a function of the overlap for the F-model for $P = 0.05$ at $T_d \simeq 0.6608$ 
(dashed line) and $T_K \simeq 0.6607$ (continuous line). Note the scale of the free-energy.}
\label{figF}
\end{figure}

As for the previous model, we have first obtained the development in powers of $Q^{\alpha \beta}$ of the Gibbs
free energy for the completely connected model, corresponding to $d=\infty$, by using the scaling $\Delta J^2=1/N$:
\begin{equation}\label{F_MF_replicated}
\begin{split}
&-\beta A\simeq \\& N\Bigg[\Biggr.
\sum_{\alpha\neq\beta}-\frac{3t}{4}(Q^{\alpha\beta})^2+
\frac{w_1}{2}\tr(Q^3)+\sum_{\alpha\neq\beta}\frac{w_2}{2}(Q^{\alpha\beta})^3+...\Bigg] \ ,
\end{split}
\end{equation}
where $\langle S_i^{a,\alpha} S_i^{a,\beta} \rangle=Q^{\alpha \beta}$ independently of $a$, \textit{i.e.} the symmetry 
between the spins is not broken. The coefficients of the expansion are equal to
\[
t=1-\beta^2 \langle (S_1)^2 \rangle \,, \,
w_1= \langle (S_1)^2 \rangle^3 \,,\,  w_2=\langle S_1 S_2 S_3 \rangle^2
\]
where $\langle (S_1)^2 \rangle$ and $\langle S_1 S_2 S_3 \rangle$ are on-site spin-spin correlation functions that 
can be easily computed: see appendix D. Because of the  property of the probability measure $\mu$ that we have chosen, 
it is easy to show that $\langle (S_1)^2 \rangle=\langle S_1 S_2 S_3 \rangle$. 
Then, by using that $\langle (S_1)^2 \rangle<1$ (the equality only applies to the $P=1$ case that we disregard),  
we obtain that $w_2>w_1$. In consequence, this model, as the previous ones, undergoes a discontinuous glass transition 
(RFOT) at a temperature higher than $T_{local}$ (always defined as the temperature at which the coefficient $t$ becomes
negative).

We have also worked out the complete mean-field solution of the model and found that if $P$ is too small the glass 
transition is pre-empted by a standard first-order transition from the high-temperature paramagnetic (liquid) state 
directly to the amorphous glass state. For values of $P$ larger than $P\simeq 0.035$ the model displays a RFOT which 
becomes less discontinuous as P is increased (the barrier at the transition decreases).
For instance, for $P = 0.05$ we find $T_d \simeq 0.6608$ and $T_K \simeq 0.6607$ and $Q(T_K)\simeq 0.073$. 
As found for the $M$-$p$-model, the Gibbs free energy  is very small for the whole range of overlap $Q$ between the first and 
the second minimum: see Fig. \ref{figF}. This feature is observed for any choice of $P$ and for all variants of the weights 
that we have analyzed. It suggests, as for the $M$-$p$-model, that the predictions of the infinite-dimensional 
mean-field theory remain valid for very high dimensions only.

In addition, we have indeed solved the model on a random graph  
of connectivity $c=6$ by using the cavity method. As anticipated, we only find only a continuous 
spin-glass transition as the temperature is lowered.
\subsection{Surface tension and a possible origin for the lack of robustness of infinite-dimensional mean-field glassy 
phenomenology}

In the infinite dimensional versions of the $M-p$-spin and F-models, we have found that the typical scale of the 
Gibbs free energy between the first and the second minimum is very small, much smaller than $T_K$. In the following, we 
argue along lines similar to those in Ref.~[\onlinecite{eastwood02}] that this makes the mean-field predictions very fragile, 
except in very high dimensions.

The Gibbs free energy as a function of the overlap is the starting point to compute the so-called ``surface tension'' $Y$; 
the latter is equal to the reduction, per unit area, of the total configurational entropy that is caused by fixing the overlap 
at the value of the secondary minimum outside a ball of radius $R$. This quantity plays a crucial role in establishing that the 
glass transition  in finite dimension is associated to a growing length-scale, called ``point-to-set correlation length'', 
and a growing time-scale \cite{wolynesbook}. Moreover, it is also the starting point for RG analyses of the glass 
transition\cite{CBTT,CammarotaPNAS}. Studies performed by using instanton techniques~\cite{dzero,silvio} have shown that the first 
non-perturbative corrections to mean-field theory lead to a value of $Y$ equal to $\int_0^{Q_{EA}}dQ \sqrt{\beta A/(m-1)}$. 
Thus, a Gibbs free energy that is very small between the two minima $Q=0$ and $Q=Q_{EA}$, as we found for the previous models, 
leads to $Y\ll T_K$.
This is problematic because, as it has become clear in recent years, some of the fluctuations not considered within mean-field 
theory correspond to adding an effective quenched disorder and lead to local fluctuations of the surface tension $Y$ and of the 
configurational entropy density $s_c$\cite{wolynes,parisi,futurepreprint}. Actually, this is expected on intuitive grounds: 
on small length scales, the values of $Y$ and $s_c$ must fluctuate in an amorphous system, especially that envisioned as 
a ``mosaic state''\cite{KTW}. The effective quenched disorder affecting $Y$ on microscopic length scales is expected to be of the order of the 
typical energy scale, {\it i.e.} $T_K$,  since it is due to local microscopic fluctuations. In consequence, on small length 
scales the disorder is much larger than the surface tension $Y$ when the latter is much smaller than $T_K$. In this case, 
heuristic arguments, as well as RG-based ones\cite{CammarotaPNAS}, suggest that the RFOT found within mean-field theory is 
destroyed. This is similar to what happens for the random-field Ising model in finite dimension if the disorder is too strong compared to the 
ferromagnetic coupling (which plays the role of the surface tension $Y$). 

The same phenomenon is present in the Potts glass model for a reasonable number (less than $10$) of states. Eastwood and 
Wolynes\cite{eastwood02} have argued that one could cure the problem and make the surface tension $Y$ large by considering 
a large number of states, $q\simeq1000$, thereby preserving the RFOT in three dimensions. However, 
this is not a proper resolution, as one then encounters another obstacle: as we have previously discussed, one indeed 
expects that frustration, and as a result glassiness, disappear for large $q$ in three dimensions. 
 
\section{Fragility of the RFOT scenario  for finite dimensional systems and proper effective theory}   
   
As already stressed, there are two main issues to address in order to understand the effect 
of fluctuations on the RFOT: 
(1) Once short-range fluctuations are taken into account, is the resulting effective theory RFOT-like? To be so, the model should favor 
two phases: one (the liquid) in which replica remain uncorrelated and another (the ideal glass) in which replica are correlated and display a high 
overlap between typical inter-replica configurations (the latter phase being metastable with respect to the former for $T>T_K$).  
(2) If the answer to the previous question is affirmative, do long-distance, possibly non-perturbative, fluctuations 
destroy the transition and alter the scenario developed by mean-field and heuristic arguments or not?

These two issues are {\it mutatis mutandis} always present in physics but the first one is often of little relevance and easily 
handled in the field of critical phenomena and phase transitions. This is not so in the present case. We have found in the case 
of the disordered spin models under study a striking fragility of the RFOT scenario to the effect of short-range fluctuations. 
Despite the fact that the description obtained from the infinite-dimensional/fully connected limit displays a ubiquitous RFOT, the models instead 
show in finite dimensions no glassy behavior at all or one that is characteristic of a conventional spin-glass transition. The answer to the above 
first question is therefore negative for these models. 
This leads us to propose that a model with a putative RFOT should better be 
first tested for the effect of short-range fluctuations only, before carrying out a full-blown calculation 
including fluctuations on all scales either by computer simulation or RG treatment.
This makes sense for a schematic model 
that is itself a caricature of the glass-forming liquids one is aiming at describing. For a realistic glass-forming liquid model, the question is rather 
whether the RFOT approach has any chance to describe its glassy behavior. The issue is then whether one can arrive, by some reasonably 
well-defined  coarse-graining or RG procedure that integrates the short-range fluctuations,  at an effective theory in the expected universality 
class of the RFOT. We discuss below these two aspects without dwelling too much on the first one which we have already addressed.

{\it (1) Does a given microscopic schematic model lead to RFOT-like behavior once short range fluctuations are included?}

Most schematic models showing a RFOT in the infinite dimensional mean-field limit that have been proposed are formulated in terms 
of discrete variables on a lattice. We suggest to use the Bethe approximation as a first test of the influence of introducing a finite 
connectivity in the problem. The Bethe approximation is equivalent to solving the model on random regular graphs. These have the same 
finite connectivity as the Euclidean lattices that they mimic, but of course not the same behavior at long distance due to the absence of loops 
on all scales and the essentially tree-like structure. Thus, within the Bethe approximation, 
one is able to take into account at least some short-range fluctuations (one can always improve the results by cluster methods) and as a result 
get a reasonable description of the local physics. If short-range fluctuations completely change the behavior of a model, one expects the Bethe 
approximation to be able to capture this effect.

As an example, we mention again the $10$-state Potts glass studied by Brangian et al.\cite{brangian02}. 
The fact that no glass transition was found in the computer simulation study on the three-dimensional cubic lattice could in 
retrospect have been predicted from the result of the Bethe approximation which already shows the disappearance of any 
RFOT for the same connectivity. The approximation is known to overlook the effect of long-distance fluctuations and is mean-field in character. 
However, it does have some merit to study the physics at short distance.   

{\it (2) Is the replicated Ginzburg-Landau free energy predicting a strong RFOT (at the mean-field level) representative of 
realistic glass-forming liquid models?} 

The Ginzburg-Landau functional that has been postulated to represent the effective theory of structural glasses\cite{dzero} 
is given in a replica formalism by
\begin{equation}
\label{effective_replicaGL}
\begin{aligned}
\beta \mathcal{F}[Q] = &\int d^d x \bigg (\sum_{\alpha\neq\beta}\Bigg[\frac{1}{2}(\partial Q^{\alpha\beta}(x))^2 
+ \frac{t}{2}Q^{\alpha\beta}(x)^2 \\& - \frac{w_2}{6}Q^{\alpha\beta}(x)^3 + \cdots \bigg] - \frac{w_1}{6}\tr(Q(x)^3)\Bigg )
\end{aligned}
\end{equation}
where $Q^{\alpha\beta}(x)$ is a local, but coarse-grained, (matrix) order parameter describing the overlap among replicas and the  
number $m$ of replicas has to be taken to $1$, $m\rightarrow 1^+$, to describe the physics associated with an exponentially large number 
of metastable states (above the RFOT). 
At the mean-field level, the glass transition can be made as strongly discontinuous as wanted by increasing $w_2-w_1$. (On the other hand, 
the transition becomes that of a conventional spin glass in a magnetic field when $w_2<w_1$.)

However, besides heuristic arguments about coarse-graining\cite{stevenson} and guidance from the 
infinite-dimensional/fully-connected limit [see \textit{e.g.} Eqs. (\ref{effectiveF}) and (\ref{F_MF_replicated})], 
there has been no serious derivation of the above functional from a proper renormalization 
step accounting for the contribution of the local, short-range fluctuations in a realistic model of glass-forming liquid. As we have precisely seen 
above how fragile  to the latter fluctuations is the RFOT scenario in the case of disordered spin models, this step should be a prerequisite to 
validate the RFOT theory as a starting point for describing the glass transition in real systems. Long-range fluctuations can of course destabilize 
the RFOT even when starting from the replicated Ginzburg-Landau functional with a strongly discontinuous RFOT at the mean-field level, but 
matters would be much worse if the RFOT is already wiped out by short-range fluctuations. The latter possibility is  advocated by Moore and 
coworkers\cite{moore} who suggest that coarse-graining a glass-forming liquid model leads to a Ginzburg-Landau functional akin to that in 
Eq. (\ref{effective_replicaGL}), but with $w_2<w_1$, hence with no RFOT at all.

A possible means to capture short-range fluctuations in liquid models is to consider the system within a cavity with amorphous boundary conditions. 
This method has recently been proposed \cite{BB} and applied \cite{PS} to measure the growth of amorphous order in glassy systems. It amounts 
to studying the thermodynamics of a system constrained to have all particles (or spins) outside a cavity in the same positions (or configuration) 
as those of a typical equilibrium configuration. By construction, degrees of freedom are then integrated out only up to a length-scale which is  the 
size of the cavity.  Provided that the size of the cavity is not too large, for instance by working far enough from the putative RFOT so that the 
point-to-set correlation length is not much bigger than the size of the atoms (which in practice is always the case in computer simulations), 
such a numerical experiment may then provide the information we are after: if the crossover between a high overlap with the reference configuration 
for a small cavity size to a small overlap for a large cavity size becomes sharper and takes place at an increasing length scale as one decreases 
the temperature, this is the sign that on the probed length scale, the finite-size system behaves according to the predicted RFOT scenario. It 
may still be hard to go from here to a determination of the parameters in the effective Ginzburg-Landau functional but it is enough 
to dismiss the alternative scenario with $w_2<w_1$ and a (possibly avoided) continuous spin-glass transition\cite{moore}. In the latter case, 
the system constrained in a cavity would display a completely different behavior. In particular, in the first regime, when the point to set starts to grow,
 the overlap would always smoothly decrease as a function of the cavity size, without leading to a sharp crossover, and this behavior would be observed on a small, never increasing length scale for overlaps 
of the order of one. Only the tail at large cavity size would display a decrease to zero, characterized by a range growing as the 
spin-glass correlation length: this however is besides our goal of probing only the influence 
of the local fluctuations \cite{footnote3b}.
 The published results of computer simulations of glass-forming liquids with the cavity method\cite{PS}  point in favor of 
the RFOT scenario, giving credit to the effective RFOT functional as a \textit{bona fide} starting point. Again, we reiterate that passing the test 
concerning the robustness with respect to the short-range fluctuations still does not guarantee the fate of the scenario at long distance when 
fluctuations on all scales are taken into account.

\section{Conclusion}
\label{conclusion}

In order to clarify the starting point for studying the effect of the fluctuations on the discontinuous glass 
transition (RFOT) found in infinite-dimensional (fully connected) glass models, we have analyzed disordered spin models 
characterized by interactions that couple spins on pairs of nearest-neighbor sites on a lattice. (This 
restriction to short-range two-site interactions was motivated by the fact that such models are amenable 
to real-space RG analyses and that they are easier to study numerically than multi-site interaction systems 
akin to the $p$-spin model.) 

Proceeding along the way, we have disentangled two possible causes for destabilizing an RFOT in finite 
dimensions. The more obvious, which was anticipated, is that long-distance, possibly non-perturbative, fluctuations 
could wipe out, or alter, the transition. The other one is that already accounting for short-range fluctuations, as the local effect of  
a finite number of neighbors, could drastically change the physics of a finite-dimensional system compared to its
infinite-dimensional counterpart. Our present work is an assessment of the latter, which shows the fragility of the RFOT 
scenario to the effect of short-range fluctuations in the case of disordered spin models. This highlights how nontrivial is 
the step of deriving an effective theory for the RFOT phenomenology from a rigorous integration of the local, short-range fluctuations.

We have identified and studied several mechanisms hampering the existence of a RFOT in finite (not too high) 
dimensional disordered spin models. For Potts glasses, the two-fold requirement of 
having (1) mainly antiferromagnetic couplings in order to avoid long-range ferromagnetic ordering and (2) a 
large number of states (or colors) in order to have a strongly discontinuous glass transition imposes that the 
lattice dimensionality has to be high, typically larger than the number of states. Otherwise, the model becomes 
unfrustrated and glassiness is wiped out. For instance, we have verified by using the Bethe approximation that 
the 10-states disordered Potts model studied by Brangian et al.\cite{brangian02} is not  
glassy in three dimensions. The other disordered models that we have considered, the $M-p$ spin 
models\cite{parisipicco} with $M=p=3$, and the F-model introduced by us, which both display a RFOT in the 
infinite-dimensional limit, are characterized by a very small surface tension $Y$. As a consequence, we expect 
finite-dimensional fluctuations to be overwhelming compared to $Y$ and to lead, in a RG sense, to a vanishing 
renormalized surface tension on larger length scales and, accordingly, to the absence of a glass transition. These 
results and the $1/d$ expansion performed for Potts glasses suggest that the perturbative regime in $1/d$ where 
(infinite-dimensional) mean-field results remain  predictive is restricted to very high dimensions only.

We stress that we have nonetheless not found a fundamental principle or a general mechanism forbidding 
{\it a priori} the existence of a RFOT and of the associated glassy phenomenology in 
all possible disordered spin models involving only interactions between pairs of nearest-neighbor sites 
in realistic dimensions (\textit{e.g.}, $d=3$). However, it is a fact---actually a puzzling one---that none of these 
models so far proposed display a RFOT beyond the large-dimensional limit. The RFOT theory originated from the 
analysis of disordered spin models. Somewhat ironically, structural glass physics turns out to lack robustness 
precisely in these models, whereas  it appears instead to be less fragile  at least to the effect of short-range fluctuations in liquid 
models\cite{footnote4}. Other simple statistical mechanical models for structural glasses have also been proposed, that 
usually consider interactions which couple degrees of freedom on more than two sites, such as plaquette models 
\cite{plaquette} and variants  (see also Refs. [\onlinecite{jorge,sasa,marinari,sarlat}] for recent alternative proposals).  Models with multi-site interactions 
and no quenched disorder such as lattice-glass models do show the anticipated glassy behavior\cite{birolimezard,tarzialatticeglass,
birolireichman} but they are prone to simple long-range ordering, in particular to crystallization. The addition of quenched 
disorder could then possibly prevent the formation of the crystalline phase without altering too much the glassy one\cite{karmakar}. In any case, 
we suggest that any new proposal should be first tested to assess the robustness of the RFOT to the effect of short-range 
fluctuations, by {\it e.g.} studying the Bethe approximation or similar procedures. The quest for a simple statistical 
mechanical model showing RFOT behavior is still open.  

{\bf Acknowledgements} 

We thank F. Caltagirone, U. Ferrari, F. Krzakala, L. Leuzzi, E. Marinari, M. Moore, G. Parisi and T. Rizzo for discussions. GB, CC and MT acknowledges financial support from the ERC grant NPRG-GLASS.   

\appendix

\section{Proof of the absence of phase transition in the  antiferromagnetic Potts model with random 
color permutations on Euclidean lattices}
\label{dobrushin}

In this appendix we aim at proving, in the case of Euclidean lattices, the absence of phase transition in the antiferromagnetic 
$q$-state Potts model with random color permutations when the number of colors/states is large enough compared to the lattice 
coordination number. To do so, we generalize the the proof given by Salas and Sokal\cite{salas97} of the absence of phase transition 
for the antiferromagnetic Potts model and apply it to the disordered model with random color permutations (note that the Salas-Sokal 
proof also applies to a random-bond Potts model, provided all couplings are negative, \textit{i.e.} anti-ferromagnetic). The Salas-Sokal 
demonstration makes use of the Dobrushin-Lanford-Ruelle approach to the equilibrium statistical mechanics of infinite volume classical 
lattice systems\cite{ruelle} and of Dobrushin's uniqueness theorem\cite{dobrushin68}. They show that the antiferromagnetic $q$-color 
Potts model on a lattice of maximum coordination number $c$ has no  phase transition (at any 
temperature $T\geq 0$) when $q>2c$. Again a large number of colors/states precludes the existence of a transition.

The antiferromagnetic Potts model with random permutations on a $d$-dimensional Euclidean lattice is defined by the following hamiltonian
\begin{equation}
\label{eq_hamiltonian}
\begin{aligned}
H= J \sum_{<i,j>}\delta_{\sigma_i, \pi_{ij}(\sigma_j)},
\end{aligned}
\end{equation}
where $J>0$, $<i,j>$ denotes distinct pairs of nearest neighbors on the lattice, the spin variable $\sigma$ can take $q$ values $\{1,2,\cdots, q\}$, 
$\delta_{\sigma,\sigma'}$ is the Kronecker symbol,  and $\pi_{ij}$ is a random permutation of the $q$ colors that is attached to the (oriented) 
edge between $i$ and $j$.  This model has a ``gauge invariance'' that prevents antiferromagnetic ordering\cite{marinari99,randompermutationnumerics,krzakala07}: indeed, if one permutes the states of the spin at a given site $i$, one can always find random permutations associated with all  edges emanating from $i$ such that the energy does not change; after averaging over the quenched disorder 
(random permutations), the (staggered) magnetization is then zero.

Following the Salas-Sokal development, we introduce the conditional probability distribution for a single spin $\sigma_i$ at site $i$ with 
external conditions given by all other spins $\{\sigma_k\}'$ with $k\neq i$. The new ingredient is that this probability distribution is defined 
for a given realization $\boldsymbol{\pi}$ of the random permutations on all (oriented) edges of the infinite-volume lattice. It is given by the 
following Boltzmann-Gibbs measure,
\begin{equation}
\label{eq_conditional_prob}
\begin{aligned}
P_{\boldsymbol{\pi}}(\sigma_i\vert \{\sigma_k\}')=
Z_{\boldsymbol{\pi}}(\{\sigma_k\}')^{-1}\exp[-\beta J\sum_{k\neq i}\delta_{\sigma_i, \widehat \pi_{ik}(\sigma_k)}],
\end{aligned}
\end{equation}
where the $\textit{a priori}$ single-spin distributions over the colors are left implicit, $\beta=1/(k_B T)$, and $\widehat\pi_{ik}$ is equal to 
$\pi_{ik}$ if the edge is oriented from $i$ to $k$ and to $\pi_{ki}^{-1}$ otherwise [we have used the fact that 
$\delta_{\sigma_k, \pi_{ki}(\sigma_i)}=\delta_{\sigma_i, \pi_{ki}^{-1}(\sigma_k)}$]. Note that, as the notations $\pi$ is used for  the 
random permutations, we have chosen $P$ for the conditional probabilities, which is then different from the Salas-Sokal 
notation\cite{salas97}.

We also introduce the quantity $c^{\boldsymbol{\pi}}_{ij}$ that measures the strength of the direct dependence between the spins 
on sites $i$ and $j$:
\begin{equation}
\label{eq_definition_strength}
\begin{aligned}
c^{\boldsymbol{\pi}}_{ij}\equiv \rm{sup}_{\{\sigma\},\{\widetilde\sigma\}: \sigma_k=\widetilde\sigma_k \forall k\neq j} \, 
d[P_{\boldsymbol{\pi}}(\sigma_i\vert \{\sigma\}),P_{\boldsymbol{\pi}}(\sigma_i\vert \{\widetilde \sigma\})],
\end{aligned}
\end{equation}
where $d[,]$ is the distance between the conditional probability measures whose definition is irrelevant here\cite{salas97}. It is 
easy to check that $c^{\boldsymbol{\pi}}_{ij}=0$ when $i=j$ or when $i$ and $j$ are not nearest neighbors on the lattice. 
The interesting situation is therefore when $j$ is one of the $c$ nearest neighbors of $i$, which will be assumed from now 
on (we consider here a lattice of constant coordination number $c$ but the reasoning works as well when $c$ is the 
maximum coordination number).

Salas and Sokal\cite{salas97} have proven a lemma which implies that 
\begin{equation}
\label{eq_lemma}
\begin{aligned}
c^{\boldsymbol{\pi}}_{ij}\leq \rm{max}\left [\frac{\rho_i^{\boldsymbol{\pi}}[1-f_{ij}] }{\rho_i^{\boldsymbol{\pi}}[f_{ij}]},
\frac{\rho_i^{\boldsymbol{\pi}}[1-\tilde f_{ij}] }{\rho_i^{\boldsymbol{\pi}}[\tilde f_{ij}]} \right ],
\end{aligned}
\end{equation}
where the functions $f_{ij}$ and $\tilde f_{ij}$ are defined as
\begin{equation}
\label{eq_definition_f_tildef}
\begin{aligned}
&f_{ij}(\sigma_i\vert \sigma_j)=\exp[-\beta J\delta_{\sigma_i, \widehat \pi_{ij}(\sigma_j)}]\, ,\\&
\tilde f_{ij}(\sigma_i\vert \widetilde \sigma_j)=\exp [-\beta  J\delta_{\sigma_i, \widehat \pi_{ij}(\widetilde\sigma_j)}]\,,
\end{aligned}
\end{equation}
where $\widehat \pi_{ij}$ is defined from $\pi_{ij}$ as above; $\rho_i^{\boldsymbol{\pi}}$ is the probability distribution at site 
$i$ in the presence of all its nearest neighbors except $j$,
\begin{equation}
\label{eq_definition_rho}
\begin{aligned}
\rho_i^{\boldsymbol{\pi}}(\sigma_i)=
(Z_i^{\boldsymbol{\pi}})^{-1}\, \exp[-\beta J\sum_{k/i ,k\neq j}\delta_{\sigma_i, \widehat \pi_{ik}(\sigma_k)}],
\end{aligned}
\end{equation}
where the notation $k/i$ means that $k$ is one of the $c$ nearest neighbors of $i$ and the normalization factor $Z_i^{\boldsymbol{\pi}}$ 
is the trace over the $q$ distinct states/colors that the spin $\sigma_i$ can take. Finally, $\rho_i^{\boldsymbol{\pi}}[f_{ij}]$ and 
$\rho_i^{\boldsymbol{\pi}}[f_{ij}]$ are short-hand notations for
\begin{equation}
\label{eq_definition_rho(1-f)}
\rho_i^{\boldsymbol{\pi}}[1-f_{ij}]= \sum_{\sigma_i=1}^q \rho_i^{\boldsymbol{\pi}}(\sigma_i) [1-f_{ij}(\sigma_i\vert \widetilde \sigma_j)],
\end{equation}
\begin{equation}
\label{eq_definition_rho(f)}
\rho_i^{\boldsymbol{\pi}}[f_{ij}]= \sum_{\sigma_i=1}^q \rho_i^{\boldsymbol{\pi}}(\sigma_i) f_{ij}(\sigma_i\vert \widetilde \sigma_j).
\end{equation}
Similar expressions hold for  $\rho_i^{\boldsymbol{\pi}}[1-\tilde f_{ij}]$ and $\rho_i^{\boldsymbol{\pi}}[\tilde f_{ij}]$. The demonstration 
of the Salas-Sokal lemma requires that $f_{ij}$ and $\tilde f_{ij}$ be in the interval $[0,1]$, which is verified when all couplings are 
antiferromagnetic as considered here.

Note that by construction $\rho_i^{\boldsymbol{\pi}}[1-f_{ij}]+\rho_i^{\boldsymbol{\pi}}[f_{ij}]=1$, so that 
\begin{equation}
\label{eq_ratio}
\begin{aligned}
\frac{\rho_i^{\boldsymbol{\pi}}[1-f_{ij}] }{\rho_i^{\boldsymbol{\pi}}[f_{ij}]}=\frac{1}{\frac{1}{\rho_i^{\boldsymbol{\pi}}[1-f_{ij}] }-1},
\end{aligned}
\end{equation}
and similarly for the expression with $\tilde f_{ij}$.

The procedure now involves deriving an upper bound for $\rho_i^{\boldsymbol{\pi}}[1-f_{ij}]$ and $\rho_i^{\boldsymbol{\pi}}[1-\tilde f_{ij}]$. 
From Eqs. (\ref{eq_lemma}) and (\ref{eq_ratio}) this provides then an upper bound for $c^{\boldsymbol{\pi}}_{ij}$.

An upper bound for $\rho_i^{\boldsymbol{\pi}}[1-f_{ij}]$ is derived by noting first that from the definition of $f_{ij}$ in Eq. (\ref{eq_definition_f_tildef}), 
the only nonzero contribution to the sum in Eq. (\ref{eq_definition_rho(1-f)}) comes from states such that $\sigma_i= \widehat \pi_{ij}(\sigma_j)$. 
There is only one such state. The next step is to find a bound for the contribution of any of the $q$ states $\sigma_i$ to the 
conditional probability distribution $\rho_i^{\boldsymbol{\pi}}$. Since $J>0$, the maximum possible weight is when 
$\sum_{k/i ,k\neq j}\delta_{\sigma_i, \widehat \pi_{ik}(\sigma_k)}=0$. This occurs when $\sigma_i$ is different from all 
$\widehat \pi_{ik}(\sigma_k)$'s with $k$ any nearest neighbor of $i$ different from $j$. Since there are $c-1$ such nearest 
neighbors, there are at most $q-(c-1)$ states with maximum weight. (The number of states is less when in the set 
$\{\widehat \pi_{ik}(\sigma_k)\}$ several colors are repeated.) The contribution o!
 f all the other states is strictly smaller. As a result the weight of any of the $q$ states $\sigma_i$ with the conditional probability 
 distribution $\rho_i^{\boldsymbol{\pi}}$ is always less than $1/[q-(c-1)]$. Putting together the above results we arrive at the 
 conclusion that  $\rho_i^{\boldsymbol{\pi}}[1-f_{ij}]\leq 1/[q-(c-1)]$. From Eq. (\ref{eq_ratio}) it then follows that
\begin{equation}
\label{eq_upper-bound_ratio}
\begin{aligned}
\frac{\rho_i^{\boldsymbol{\pi}}[1-f_{ij}] }{\rho_i^{\boldsymbol{\pi}}[f_{ij}]}\leq \frac{1}{q-c},
\end{aligned}
\end{equation}
and similarly for the ratio involving $\tilde f_{ij}$.

One then concludes that for all pairs of nearest neighbors $i,j$ on the lattice
\begin{equation}
\label{eq_upper-bound_c}
\begin{aligned}
c_{ij}^{\boldsymbol{\pi}}\leq \frac{1}{q-c}.
\end{aligned}
\end{equation}

For any given realization of the quenched disorder, the so-called Dobrushin constant $\alpha^{\boldsymbol{\pi}} 
\equiv \rm{sup}_i \sum_{j\neq i} c_{ij}^{\boldsymbol{\pi}}$ therefore satisfies
\begin{equation}
\label{eq_upper-bound_alpha}
\begin{aligned}
\alpha^{\boldsymbol{\pi}}  \leq \frac{c}{q-c}.
\end{aligned}
\end{equation}
The Dobrushin uniqueness theorem\cite{dobrushin68} states that there is a unique infinite-volume Gibbs measure when 
the condition $\alpha^{\boldsymbol{\pi}}<1$ is satisfied. Under the same condition, an additional theorem implies that the 
correlations in the infinite-volume Gibbs measure decay exponentially with distance (see Ref. [\onlinecite{salas97}] and 
references therein). By using Eq. (\ref{eq_upper-bound_alpha}), the above condition amounts to
\begin{equation}
\label{eq_uniqueness}
\begin{aligned}
q > 2c.
\end{aligned}
\end{equation}
This is true for any realization of the random permutations on the infinite-volume lattice. This proves the absence of  phase transition 
and the exponential decay of the correlations, hence the absence of any glassy behavior, when the number of colors is larger than a 
threshold depending on the lattice coordination number. 

\section{The $1/d$ expansion for the Potts model}
\label{1overd}

As discussed in the main text, we use the simplex representation of the Potts glass:
\begin{equation}
H=-\sum_{<i,j>}J_{ij} \sum_{a=1}^{q-1} S_{i,a}S_{j,a} \ .
\end{equation}
In this representation $q$ is the number of colors and the degrees of freedom $S_{i,a}$ are vectors 
pointing toward the $q$ vertices of a thetrahedron in a  $(q-1)$-dimensional space.
These vectors satisfy the following relations:
\begin{equation}
\sum_{s=1}^q e^s_a e^s_b=q\, \delta_{a b}
\label{p1}
\end{equation}
\begin{equation}
\sum_{a=1}^{q-1} e^s_a e^{s'}_a=q\, \delta_{s s'}-1
\label{p2}
\end{equation}
\begin{equation}
\sum_{s=1}^q e^s_a=0
\label{p3}
\end{equation}
\begin{equation}
\left(\sum_{a=1}^{q-1}e^s_a e^{s'}_a\right)^2=(q-2)\sum_{a=1}^{q-1} e^s_a e^{s'}_a+q-1 \ .
\label{p4}
\end{equation}
Integration of the partition function replicated $n$ times over the quenched disorder gives:
\begin{eqnarray}
\overline{Z^n}&=&\sum_{\{s^{\alpha}\}}\prod_{<i,j>}\exp\left[\frac{\beta^2}{2d}\sum_{(\alpha,\beta)}
\sum_{a,b} S^{\alpha}_{i,a}S^{\beta}_{i,b}S^{\alpha}_{j,a}S^{\beta}_{j,b}\right. \nonumber\\
&+& \left.\beta\tilde{J}_0\sum_{\alpha}\sum_a S^{\alpha}_{i,a}S^{\alpha}_{j,a}\right] \ ,
\end{eqnarray}

\subsection{Large-dimension (or large-temperature) expansion}

The large-dimension expansion is a kind of high-temperature expansion, as discussed by Georges and Yedidia \cite{yedidia2}. 
Following Refs. [\onlinecite{yedidia1,yedidia2}] we introduce a small parameter $\upsilon$ in front of the terms that couples different sites.
We also introduce two sets of Lagrange multipliers $\lambda^{\alpha\beta}_i$ and $\eta_i^{\alpha}$ to ensure that the following 
relations hold for each value of the small parameter $\upsilon$:
\begin{equation} 
\lrg{S^{\alpha}_{i,a} S^{\beta}_{i,b}}=\delta_{ab}\, Q_i^{\alpha\beta} 
\label{q}
\end{equation}
and
\begin{equation}
\lrg{S^{\alpha}_{i,a}}=m_i^{\alpha} \ .
\label{m}
\end{equation}
As a result, we have
\begin{equation}
\begin{split}
\label{-bA}
&-\beta A(\{Q_i^{\alpha\beta}\},\{m_i^{\alpha}\};\{\lambda_i^{\alpha\beta}\},\{\eta_i^{\alpha}\})=\ln \tr_{\{S^{\alpha}\}} 
\exp \bigg[ \upsilon H \\&
+ \sum_i\sum_{\alpha,\beta} \lambda_i^{\alpha\beta}\left(\sum_aS^{\alpha}_{i,a}S^{\beta}_{i,a}-
(q-1)q_i^{\alpha\beta}\right) \\&
+ \sum_i\sum_{\alpha}\eta_i^{\alpha}\left(\sum_aS_{i,a}^{\alpha}-(q-1)m_i^{\alpha}\right)\bigg]
\end{split}
\end{equation}
where $N$ is the number of sites in the system and 
\begin{equation}
\begin{split}
H=&\frac{\beta^2}{2d} \sum_{(i,j)} \sum_{(\alpha,\beta)} \sum_{a,b} S^{\alpha}_{i,a} S^{\beta}_{i,b} 
S^{\alpha}_{j,a} S^{\beta}_{j,b} \\&+ \beta \tilde{J}_0\sum_{(i,j)}\sum_{\alpha}\sum_a S^{\alpha}_{i,a}S^{\alpha}_{j,a} \ .\nonumber
\end{split}
\end{equation}
We have implicitly assumed that by symmetry 
$$Q_{i,ab}^{\alpha\beta}=\delta_{ab}Q^{\alpha\beta}_i \ \ \forall a$$
$$m_{i,a}^{\alpha\beta}=m^{\alpha}_i \ \ \forall a \ .$$
To preserve the relations in Eqs. \eqref{q} and \eqref{m} we impose $\lambda_i^{\alpha\beta}$ so that
\begin{equation}
\frac{\partial A}{\partial \lambda_i^{\alpha\beta}}=0 \ \ \mbox{, which gives} \ \ \ Q_i^{\alpha\beta}=\frac{1}{q-1}\sum_aS^{\alpha}_{i,a} S^{\beta}_{i,a}
\end{equation}
and $\eta_i^{\alpha}$ so that
\begin{equation}
\frac{\partial A}{\partial \eta_i^{\alpha}}=0 \ \ \mbox{, which gives} \ \ \ m_i^{\alpha}=\frac{1}{q-1}\sum_aS^{\alpha}_{i,a} \ .
\end{equation}

A large-temperature/large-dimension expansion of $-\beta A$ can be obtained expanding 
$\exp(\upsilon H)$ around $\upsilon=0$ and putting $\upsilon=1$ at the end\cite{yedidia1}.
The terms of the expansion are expressed in terms of averages $\lrg{\;}_0$ with the same weight as in Eq. \eqref{-bA}, 
except that $\upsilon$ is set to zero:
\begin{equation}
\begin{split}
&\lrg{\mathcal{O}(\{S^{\alpha}\})}_0=
\\&
\tr_{\{S^{\alpha}\}}\left\{\exp\left[\sum_i\sum_{\alpha,\beta} \lambda_i^{\alpha\beta}\left(\sum_aS^{\alpha}_{i,a}S^{\beta}_{i,a}
-(q-1)Q_i^{\alpha\beta}\right)\right.\right.\\
&+\left.\left.\sum_i\sum_{\alpha}\eta_i^{\alpha}\left(\sum_aS_{i,a}^{\alpha}-(q-1)m_i^{\alpha}\right)\right]\mathcal{O}\right\}/Z_0
\end{split}
\end{equation}
where 
\begin{equation}
\begin{split}
&Z_0=\\&\tr_{\{S^{\alpha}\}}\exp\bigg[\sum_i\sum_{\alpha,\beta} \lambda_i^{\alpha\beta}\left(\sum_aS^{\alpha}_{i,a}S^{\beta}_{i,a}
-(q-1)Q_i^{\alpha\beta}\right)\\&
+ \sum_i\sum_{\alpha}\eta_i^{\alpha}\left(\sum_aS_{i,a}^{\alpha}-(q-1)m_i^{\alpha}\right)\bigg] \ .
\end{split}
\end{equation}

The expansion is of the form
\begin{equation}
A=A_0+A_1\upsilon+\frac{1}{2}A_2\upsilon^2+\dots \ .
\label{develop}
\end{equation}
with, at the zeroth order,
\begin{equation}
-\beta A_0=\atz{(-\beta A)}=\ln Z_0 \ ,
\end{equation}
and at the first order,
\begin{equation}
-\beta A_1=\atz{\frac{\partial (-\beta A)}{\partial \upsilon}}\,.
\end{equation}
Then,
\begin{eqnarray}
&&-\beta A_1=\lrg{H}=\frac{\beta^2}{2d} \sum_{(i,j)} \sum_{(\alpha,\beta)} \sum_{a,b} \lrg{S^{\alpha}_{i,a} S^{\beta}_{i,b} 
S^{\alpha}_{j,a} S^{\beta}_{j,b}}_0 \nonumber \\
&+& \beta \tilde{J}_0\sum_{(i,j)}\sum_{\alpha}\sum_a \lrg{S^{\alpha}_{i,a}S^{\alpha}_{j,a}}_0 
\end{eqnarray}
and by using Eq. \eqref{q} we obtain that $-\beta A_1$ is equal to
\begin{equation}
\frac{\beta^2}{2d}(q-1)\sum_{<i,j>}\sum_{(\alpha,\beta)}Q_i^{\alpha\beta}Q_j^{\alpha\beta}
+\beta \tilde{J}_0(q-1)\sum_{<i,j>}\sum_{\alpha}m_i^{\alpha}m_j^{\alpha} \ .\nonumber
\end{equation}

To more easily compute the second order we introduce the operator 
\begin{eqnarray}
&&\mathcal{U}= \nonumber \\
&& H-\lrg{H}+\sum_i\sum_{(\alpha,\beta)}\frac{\partial \lambda_i^{\alpha\beta}}{\partial \upsilon}\left(\sum_aS^{\alpha}_{i,a}
S^{\beta}_{i,a}-(q-1)Q_i^{\alpha\beta}\right)\nonumber \\
&&+\sum_i\sum_{\alpha}\frac{\partial \eta_i^{\alpha}}{\partial \upsilon}\left(\sum_aS_{i,a}^{\alpha}-(q-1)m_i^{\alpha}\right) ,
\label{Udef}
\end{eqnarray}
which is such that 
\begin{equation}
\frac{\partial\lrg{\mathcal{O}}}{\partial\upsilon}=\lrg{\frac{\partial\mathcal{O}}{\partial\upsilon}}+\lrg{\mathcal{OU}} \ .
\label{Uorigin}
\end{equation}
Useful properties of the operator $\mathcal{U}$ are listed in Ref. [\onlinecite{yedidia1}]. 
The term $A_2$ can be written as
\begin{equation}
-\beta A_2=\atz{\frac{\partial^2 (-\beta A)}{\partial \upsilon^2}}=\lrg{\mathcal{U}^2_0}_0 \ ,
\end{equation}
where $\mathcal{U}_0$ is the operator $\mathcal{U}$ such that all the averages it contains are those with $\upsilon=0$ in the weight.
Focusing on the $m=0$ case (no ferromagnetic ordering) we get
\begin{eqnarray}
&&-\frac{1}{2}\beta A_2=\sum_{<i,j>}
\label{bA2}
\\
\nonumber
&\Biggl[\Biggr.&
\dfrac{\beta^4}{8d^2}
\sum_{\supsubdue{(\alpha,\beta)}{(\gamma,\delta)}}
\sum_{\supsubdue{a,b}{c,d}}
\left(\lrg{S^{\alpha}_{i,a}S^{\beta}_{i,b}S^{\gamma}_{i,c}S^{\delta}_{i,d}}_0
-\delta_{ab}\delta_{cd}Q^{\alpha\beta}_i Q^{\gamma\delta}_i\right)\\
&&
\times \left(\lrg{S^{\alpha}_{j,a}S^{\beta}_{j,b}S^{\gamma}_{j,c}S^{\delta}_{j,d}}_0
-\delta_{ab}\delta_{cd}Q^{\alpha\beta}_j Q^{\gamma\delta}_j\right)\nonumber
\\
\nonumber
&+&\dfrac{\beta^2}{2d}\beta \tilde{J}_0
\sum_{\supsubdue{(\alpha,\beta)}{\gamma}}
\sum_{\supsubdue{a,b}{c}}
\lrg{S^{\alpha}_{i,a}S^{\beta}_{i,b}S^{\gamma}_{i,c}}_0
\lrg{S^{\alpha}_{j,a}S^{\beta}_{j,b}S^{\gamma}_{j,c}}_0
\\
\nonumber
&+&\frac{1}{2}\left(\beta \tilde{J}_0\right)^2
\sum_{\alpha\neq\beta}
(q-1)Q^{\alpha\beta}_i Q^{\alpha\beta}_j
\Biggl.\Biggr] \ ,
\end{eqnarray}
as in the last sum the terms with $\alpha=\beta$ give only a constant.

\subsection{Zeroth- and first-order terms}

The first-order term is directly obtained as
\begin{equation}
\label{bA1}
-\beta A_1=\frac{\beta^2}{2d}\sum_{<i,j>}\sum_{(\alpha,\beta)}(q-1)Q_i^{\alpha\beta}Q_j^{\alpha\beta} \ .
\end{equation}
On the other hand, the zeroth-order term reads
\begin{equation}
\begin{split}
\label{bA0}
&-\beta A_0=\\&\ln\tr_{\{S^{\alpha}\}}\exp\left[\sum_i\sum_{(\alpha,\beta)}
\lambda_i^{\alpha\beta}\left(\sum_a S^{\alpha}_{i,a}S^{\beta}_{i,a}-(q-1)Q_i^{\alpha\beta}\right)\right] \ .
\end{split}
\end{equation}
In the following we compute this expression in an expansion in $\lambda_i^{\alpha\beta}$ and $Q_i^{\alpha\beta}$. First, we pull out the 
term not involved in the trace, which gives
\begin{equation}
\begin{split}
&-\beta A_0=-\sum_i \sum_{(\alpha,\beta)}(q-1)\lambda_i^{\alpha\beta}Q_i^{\alpha\beta}\\&
+\ln\tr_{\{S^{\alpha}\}}\prod_i\exp\left(\sum_{(\alpha,\beta)}\lambda_i^{\alpha\beta}\sum_aS^{\alpha}_{i,a}S^{\beta}_{i,a}\right) \ .
\end{split}
\end{equation}
Next,  we expand the argument of the trace to cubic order in the $\lambda_i$'s:
\begin{equation}
\begin{split}
&\ln\tr_{\{S^{\alpha}\}}\prod_i\exp\Biggl(\sum_{(\alpha,\beta)}\lambda_i^{\alpha\beta}\sum_aS^{\alpha}_{i,a}S^{\beta}_{i,a}\Biggr)=
\\&
\ln\prod_i\tr_{\{S^{\alpha}_i\}}\Biggl(1+\sum_{(\alpha,\beta)}\lambda_i^{\alpha\beta}\sum_aS^{\alpha}_{i,a}S^{\beta}_{i,a}\, +\\&
\frac{1}{2}\sum_{\supsubdue{(\alpha,\beta)}{(\gamma,\delta)}}\lambda_i^{\alpha\beta}\lambda_i^{\gamma\delta}\sum_{a,c}
S^{\alpha}_{i,a}S^{\beta}_{i,a}S^{\gamma}_{i,c}S^{\delta}_{i,c}\, +\\&
\frac{1}{6}\sum_{\supsubtre{(\alpha,\beta)}{(\gamma,\delta)}{(\epsilon,\zeta)}}\lambda_i^{\alpha\beta}\lambda_i^{\gamma\delta}
\lambda_i^{\epsilon\zeta}\sum_{a,c,e}S^{\alpha}_{i,a}S^{\beta}_{i,a}S^{\gamma}_{i,c}S^{\delta}_{i,c}S^{\epsilon}_{i,e}S^{\zeta}_{i,e}\Biggr) \ .
\end{split}
\end{equation}
From Eqs. \eqref{p1} and \eqref{p3} it is easily realized that one has to pair the replicas to obtain nonzero contributions from the trace.
As a result, we find 
\begin{equation}
\begin{split}
\tr_{\{S^{\alpha}\}}\sum_{(\alpha,\beta)}\lambda^{\alpha\beta}\sum_aS^{\alpha}_aS^{\beta}_a=0 \ ,
\end{split}
\end{equation}
\begin{equation}
\begin{split}
&\frac{1}{2}\tr_{\{S^{\alpha}\}}\sum_{\supsubdue{(\alpha,\beta)}{(\gamma,\delta)}}\lambda^{\alpha\beta}\lambda^{\gamma\delta}
\sum_{a,c}S^{\alpha}_aS^{\beta}_aS^{\gamma}_cS^{\delta}_c=\\& \frac{1}{2}\sum_{(\alpha,\beta)}(\lambda^{\alpha\beta})^2(q-1)q^n \ ,
\end{split}
\end{equation}
and
\begin{eqnarray}
&&\frac{1}{6}\tr_{\{S^{\alpha}\}}\sum_{\supsubtre{(\alpha,\beta)}{(\gamma,\delta)}{(\epsilon,\zeta)}}\lambda^{\alpha\beta}
\lambda^{\gamma\delta}\lambda^{\epsilon\zeta}\sum_{a,c,e}S^{\alpha}_aS^{\beta}_aS^{\gamma}_cS^{\delta}_c
S^{\epsilon}_eS^{\zeta}_e=\nonumber\\
&&\frac{1}{6}\sum_{(\alpha,\beta)}(\lambda^{\alpha\beta})^3(q-1)(q-2)q^n+\frac{1}{6}\tr{\lambda^3}(q-1)q^n \ ,\nonumber
\end{eqnarray}
where we have used Eq. \eqref{p1} and
\begin{equation}
\sum_{a,b,c}\left(\sum_{s=1}^q e^s_a e^s_b e^s_c\right)^2=(q-1)(q-2)/q^2 \ .
\end{equation}
Finally, the expression of the zeroth-order contribution is obtained to a $O(\lambda^4)$ as
\begin{eqnarray}
\label{bA0f}
&&-\beta A_0=N(q-1)\left(-\sum_{(\alpha,\beta)}\lambda^{\alpha\beta}Q^{\alpha\beta}+n\ln q\right.\\
&&\left.+\frac{1}{2}\sum_{(\alpha,\beta)}(\lambda^{\alpha\beta})^2+\frac{q-2}{6}\sum_{(\alpha,\beta)}(\lambda^{\alpha\beta})^3
+\frac{1}{6}\tr{\lambda^3} 
\right) \ .\nonumber
\end{eqnarray}

\subsection{Second order}

We start with Eq. \eqref{bA2}. To compute this expression as a function of $\lambda_i^{\alpha\beta}$ and $Q_i^{\alpha\beta}$ we 
again expand the weight in $\lrg{\,}_0$ around $\lambda^{\alpha\beta}=0$ to a $O(\lambda^4)$. 
The only contributions that can be nonzero involve
\begin{equation}
\mathcal S_3^{\alpha \beta \gamma}=\lrg{S^{\alpha}_{i,a}S^{\beta}_{i,b}S^{\gamma}_{i,c}}_0
\end{equation}
with $\alpha\neq\beta$, for
\begin{enumerate}
\item $\gamma\neq\alpha$ and $\gamma\neq\beta$
\item $\gamma=\alpha$ and $\gamma\neq\beta$,
\end{enumerate}
and
\begin{equation}
\mathcal S_4^{\alpha \beta \gamma \delta}=\lrg{S^{\alpha}_{i,a}S^{\beta}_{i,b}S^{\gamma}_{i,c}S^{\delta}_{i,d}}_0
\end{equation}
with $\alpha\neq\beta$ and $\gamma\neq\delta$, for
\begin{enumerate}
\item $\gamma\neq\alpha$, $\delta\neq\alpha$, $\gamma\neq\beta$ and $\delta\neq\beta$
\item $\gamma=\alpha$ and $\delta\neq\beta$
\item $\gamma=\alpha$ and $\delta=\beta$.
\end{enumerate}
In the end we need an expression that is cubic in the $Q_i^{\alpha\beta}$'s.
For each term we therefore make an expansion up to the third order in $\lambda^{\alpha\beta}_i$ at most. 
The detailed computation is quite cumbersome. Here, we only present the results term by term.
 
Let us start with the first term, $\mathcal S_3^{\alpha \beta \gamma}=\lrg{S^{\alpha}_{a}S^{\beta}_{b}S^{\gamma}_{c}}_0$ with $\alpha\neq\beta$.
\begin{enumerate}
\item When $\gamma\neq\alpha$ and $\gamma\neq\beta$, the terms of order $O(1)$ and $O(\lambda)$ are zero. 
Hence, one only has terms of order $O(\lambda^2)$ that become of order $O(\lambda^4)$ when the square is taken
in the Gibbs free energy and can therefore be neglected. 
\item When $\gamma=\alpha$, the only term of order $O(1)$ is zero and we thus have to compute terms of order $O(\lambda)$ and $O(\lambda^2)$. 
We find that the linear term $\mathcal{S}_{3,\lambda}^{\alpha\beta\alpha}$ reads
\begin{equation}
q^{-1}\lambda^{\alpha\beta}\sum_S S^{\alpha}_aS^{\alpha}_bS^{\alpha}_c
\end{equation}
whereas the quadratic term $\mathcal{S}_{3,\lambda^2}^{\alpha\beta}$ reads
\begin{eqnarray}
&&\frac{1}{2}p^{-2}(\lambda^{\alpha\beta})^2\sum_{a_1,a_2}\left(\sum_S S^{\alpha}_aS^{\alpha}_cS^{\alpha}_{a_1}
S^{\alpha}_{a_2}\right)\left(\sum_S S^{\beta}_bS^{\beta}_{a_1}S^{\beta}_{a_2}\right)\nonumber+\\
&&q^{-1}\sum_{\supsubdue{\eta\neq\alpha}{\eta\neq\beta}}\lambda^{\alpha\eta}\lambda^{\eta\beta}\left(\sum_S S^{\alpha}_a
S^{\alpha}_bS^{\alpha}_c\right) \ .\nonumber
\end{eqnarray}
\end{enumerate}
Now we consider the terms from $\mathcal S_4^{\alpha \beta \gamma \delta}=\lrg{S^{\alpha}_{a}S^{\beta}_{b}S^{\gamma}_{c}
S^{\delta}_{d}}_0$ with $\alpha\neq\beta$ and $\gamma\neq\delta$.
\begin{enumerate}
\item When $\gamma\neq\alpha$, $\gamma\neq\beta$, $\delta\neq\alpha$, and $\delta\neq\beta$, the terms of order $O(1)$ and 
$O(\lambda)$ are zero and there is no need to compute the term of order $O(\lambda^2)$.
\item When $\gamma=\alpha$ and $\delta\neq\beta$, the term of order $O(1)$ is zero but not the term of order $O(\lambda)$:
\begin{equation}
\mathcal{S}_{4,\lambda}^{\alpha\beta\alpha\delta}=\lambda^{\beta\delta}\delta_{ac}\delta_{bd} \ .
\end{equation}
\item For the quadratic term, there are several contributions that will be denoted according to the way spins are grouped in the sums:
\begin{equation}
\mathcal{S}_{4,\lambda^2}^{\alpha\beta\alpha\delta}(4,2,2)=q^{-1}\lambda^{\alpha\beta}\lambda^{\alpha\delta}\Biggl(\sum_S
S^{\alpha}_aS^{\alpha}_bS^{\alpha}_cS^{\alpha}_d\Biggr)
\end{equation}
\begin{eqnarray}
&&\mathcal{S}_{4,\lambda^2}^{\alpha\beta\alpha\delta}(3,3,2)=q^{-2}\lambda^{\beta\delta}(\lambda^{\alpha\beta}+\lambda^{\alpha\delta})
\times \nonumber\\
&&\sum_{a_1}\Biggl(\sum_SS^{\alpha}_aS^{\alpha}_cS^{\alpha}_{a_1}\Biggr)\Biggl(\sum_SS^{\beta}_bS^{\beta}_dS^{\beta}_{a_1}\Biggr)
\end{eqnarray}
\begin{eqnarray}
&&\mathcal{S}_{4,\lambda^2}^{\alpha\beta\alpha\delta}(2,3,3)=\frac{1}{2}q^{-2}(\lambda^{\beta\delta})^2\delta_{ac}\times\nonumber\\
&&\sum_{a_1,a_2}\Biggl(\sum_SS^{\beta}_bS^{\beta}_{a_1}S^{\beta}_{a_2}\Biggr)\Biggl(\sum_SS^{\delta}_dS^{\delta}_{a_1}S^{\delta}_{a_2}\Biggr)
\end{eqnarray}
\begin{equation}
\mathcal{S}_{4,\lambda^2}^{\alpha\beta\alpha\delta}(2,2,2,2)=\sum_{\supsubtre{\eta\neq\alpha}{\eta\neq\beta}{\eta\neq\delta}}
\lambda^{\beta\eta}\lambda^{\eta\delta}\delta_{ac}\delta_{bd}
\end{equation}
\item When $\gamma=\alpha$ and $\delta=\beta$, all terms are nonzero. The term of order $O(1)$ is given by
\begin{equation}
\mathcal{S}_{4,1}^{\alpha\beta\alpha\beta}=\tr_{S^{\alpha}}S^{\alpha}_aS^{\alpha}_cS^{\beta}_bS^{\beta}_d=\delta_{ac}\delta_{bd}\ .
\end{equation}
Note that it is not needed to go to the term of order $O(\lambda^3)$ because
\begin{eqnarray}
&&\lrg{\;}_0^2=(\lrg{\;}_0-\delta_{ac}\delta_{bd}+\delta_{ac}\delta_{bd})^2\\
&&=(\lrg{\;}_0-\delta_{ac}\delta_{bd})^2+2\delta_{ac}\delta_{bd}(\lrg{\;}_0-\delta_{ac}\delta_{bd})+(\delta_{ac}\delta_{bd})^2 \ .\nonumber
\end{eqnarray}
Since $\delta_{ac}\delta_{bd}\lrg{\;}_0$ is a constant, only the first term matters and computing the term $\lrg{\;}_0$ to the order 
$O(\lambda^2)$ is enough to give a result of order $O(\lambda^3)$.

The linear term reads
\begin{equation}
\mathcal{S}_{4,\lambda}^{\alpha\beta\alpha\beta}=q^{-2}\lambda^{\alpha\beta}\sum_{a_1}\Biggl(\sum_SS^{\alpha}_aS^{\alpha}_{c}
S^{\alpha}_{a_1}\Biggr)\Biggl(\sum_SS^{\beta}_bS^{\beta}_{d}S^{\beta}_{a_1}\Biggr) \ .
\end{equation}
For the quadratic one we find several contributions:
\begin{eqnarray}
&&\mathcal{S}_{4,\lambda^2}^{\alpha\beta\alpha\beta}(4,4)=\frac{1}{2}q^{-2}(\lambda^{\alpha\beta})^2\times\\
&&\sum_{a_1,a_2}\Biggl(\sum_SS^{\alpha}_aS^{\alpha}_cS^{\alpha}_{a_1}S^{\alpha}_{a_2}\Biggr)\Biggl(\sum_SS^{\beta}_bS^{\beta}_d
S^{\beta}_{a_1}S^{\beta}_{a_2}\Biggr)\nonumber
\end{eqnarray}
\begin{equation}
\mathcal{S}_{4,\lambda^2}^{\alpha\beta\alpha\beta}(4,2,2)=\frac{1}{2}(q-1)\delta_{ac}\delta_{bd}\left(\sum_{\eta\neq\alpha}
(\lambda^{\beta\eta})^2+\sum_{\eta\neq\beta}(\lambda^{\alpha\eta})^2\right)
\end{equation}
\begin{eqnarray}
&&\mathcal{S}_{4,\lambda^2}^{\alpha\beta\alpha\beta}(3,3,2)=q^{-2}\sum_{\supsubdue{\eta\neq\alpha}{\eta\neq\beta}}
\lambda^{\alpha\eta}\lambda^{\eta\beta}\times\\
&&\sum_{a_1}\Biggl(\sum_SS^{\alpha}_aS^{\alpha}_cS^{\alpha}_{a_1}\Biggr)\Biggl(\sum_SS^{\beta}_bS^{\beta}_dS^{\beta}_{a_1}\Biggr)\nonumber
\end{eqnarray}
\end{enumerate}

\subsection{The Gibbs free energy}

In Eq. \eqref{bA2} one has to consider the following terms:
\begin{equation}
{\bf S}_3=\sum_{\supsubdue{(\alpha,\beta)}{\gamma}}\sum_{\supsubdue{a,b}{c}}\lrg{S^{\alpha}_aS^{\beta}_bS^{\gamma}_c}_0^2
\label{S3}
\end{equation}
\begin{equation}
{\bf S}_4=\sum_{\supsubdue{(\alpha,\beta)}{(\gamma,\delta)}}\sum_{\supsubdue{a,b}{c,d}}\lrg{S^{\alpha}_aS^{\beta}_bS^{\gamma}_c
S^{\delta}_d}_0^2
\label{S4}
\end{equation}
\begin{equation}
{\bf T}_4=\sum_{\supsubdue{(\alpha,\beta)}{(\gamma,\delta)}}\sum_{\supsubdue{a,b}{c,d}}\delta_{ac}\delta_{bd}Q^{\alpha\beta}Q^{\gamma\delta}
\lrg{S^{\alpha}_aS^{\beta}_bS^{\gamma}_cS^{\delta}_d}_0 \ .
\label{T4}
\end{equation}
To compute these terms we use the relations obtained in Ref. [\onlinecite{cwilich}] after introducing
where 
\begin{equation}
v_{abc}=\left(\sum_SS_aS_bS_c\right) 
\end{equation}
and 
\begin{equation}
F_{abcd}=\left(\sum_SS_aS_bS_cS_d\right) 
\end{equation}
and using the Einstein convention of summing over repeated indices, one has
\begin{equation}
v_{abc}v_{ade}=qF_{bcde}-q^2\delta_{bc}\delta_{de}
\end{equation}
\begin{equation}
v_{abc}v_{abd}=q^2(q-1)\delta_{cd}
\end{equation}
\begin{equation}
v_{abc}^2=q^2(q-1)(q-2)
\end{equation}
\begin{equation}
F_{aabc}=q(q-1)\delta_{bc}
\end{equation}
\begin{equation}
F_{aabb}=q(q-1)^2
\end{equation}
\begin{equation}
F_{abcd}F_{abef}=q^2\left(\frac{q-2}{q}F_{cdef}+\delta_{cd}\delta_{ef}\right)
\end{equation}
\begin{equation}
F_{abcd}^2=q^2(q-1)(q^2-3q+3) \ ,
\end{equation}
from which it also follows that 
\begin{equation}
\delta_{ab}\delta_{cd}v_{abe}v_{cde}=0 \ .
\end{equation}

In Eq. \eqref{S3}, only the terms in which $\alpha=\gamma$ or $\beta=\gamma$ are different from zero.
This means that 
\begin{equation}
{\bf S}_3=\sum_{\alpha\neq\beta}\sum_{\supsubdue{a,b}{c}}\lrg{S^{\alpha}_aS^{\beta}_bS^{\alpha}_c}_0^2 \ .
\end{equation}
After substituting the terms of order $O(\lambda)$ and $O(\lambda^2)$, one finds
\begin{equation}
\begin{split}
&{\bf S}_3=\sum_{\alpha\neq\beta}\sum_{\supsubdue{a,b}{c}}\bigg(q^{-1}\lambda^{\alpha\beta}v_{abc} + \frac{1}{2}q^{-2}
\left(\lambda^{\alpha\beta}\right)^2\, \times \\&\sum_{a_1,a_2}F_{aca_1a_2}v_{ba_1a_2}+q^{-1}
\sum_{\supsubdue{\eta\neq\alpha}{\eta\neq\beta}}\lambda^{\alpha\eta}\lambda^{\eta\beta}v_{abc}\bigg)^2 \ ,
\end{split}
\end{equation}
which finally gives
\begin{eqnarray}
&&{\bf S}_3=(q-1)(q-2)\sum_{\alpha\neq\beta}(\lambda^{\alpha\beta})^2+\\
&&+(q-1)(q-2)^2\sum_{\alpha\neq\beta}(\lambda^{\alpha\beta})^3+
2(q-1)(q-2)\tr(\lambda^3) \ .\nonumber
\end{eqnarray}

In Eq. \eqref{S4} the nonzero terms are those for which $\gamma$ or $\delta$ is equal to $\alpha$ or $\beta$ while the other is different 
from both, or such that $\gamma$ and $\delta$ are equal to $\alpha$ and $\beta$.
Hence we have
\begin{equation}
{\bf S}_4={\bf S}_4^a+{\bf S}^b_4
\end{equation}
where 
\begin{equation}
\label{S4a}
{\bf S}_4^a=\sum_{\supsubtre{\alpha\neq\beta}{\alpha\neq\delta}{\beta\neq\delta}}\sum_{\supsubdue{a,b}{c,d}}\lrg{S^{\alpha}_aS^{\beta}_b
S^{\alpha}_cS^{\delta}_d}_0^2
\end{equation}
and 
\begin{equation}
\label{S4b}
{\bf S}^b_4=\frac{1}{2}\sum_{\alpha\neq\beta}\sum_{\supsubdue{a,b}{c,d}}\lrg{S^{\alpha}_aS^{\beta}_bS^{\alpha}_cS^{\beta}_d}_0^2 \ .
\end{equation}
Eq. \eqref{S4a} can be reexpressed as
\begin{eqnarray}
&&{\bf S}_4^a=\sum_{\supsubtre{\alpha\neq\beta}{\alpha\neq\delta}{\beta\neq\delta}}\sum_{\supsubdue{a,b}{c,d}}
\Biggl(\lambda^{\beta\delta}\delta_{ac}\delta_{bd}+
q^{-1}\lambda^{\alpha\beta}\lambda^{\alpha\delta}F_{abcd}+\nonumber\\
&&q^{-2}\lambda^{\beta\delta}(\lambda^{\alpha\beta}+\lambda^{\alpha\delta})
\sum_{a_1}v_{aca_1}v_{bda_1}+\Biggr.\\
\nonumber
&&\Biggl.
\frac{1}{2}q^{-2}(\lambda^{\beta\delta})^2\delta_{ac}
\sum_{a_1a_2}v_{ba_1a_2}v_{da_1a_2}+
\delta_{ac}\delta_{bd}\sum_{\supsubtre{\eta\neq\alpha}{\eta\neq\beta}{\eta\neq\delta}}\lambda^{\beta\eta}\lambda^{\eta\delta}\Biggr)^2 \ ,
\end{eqnarray}
which gives 
\begin{equation}
\begin{split}
{\bf S}_4^a=&-2(q-1)^2\sum_{\beta\neq\delta}(\lambda^{\beta\delta})^2
-2(q-1)^3\sum_{\beta\neq\delta}(\lambda^{\beta\delta})^3 \\&
-4(q-1)^2\tr(\lambda^3) \ .
\end{split}
\end{equation}
Eq. \eqref{S4b} can be reexpressed as
\begin{equation}
\begin{split}
&{\bf S}_4^b=\frac{1}{2}\sum_{\alpha\neq\beta}\sum_{\supsubdue{a,b}{c,d}}\Bigg(q^{-2}\lambda^{\alpha\beta}\sum_{a_1}v_{aca_1}v_{bda_1}\\&
+\frac{1}{2}q^{-2}(\lambda^{\alpha\beta})^2\sum_{a_1a_2}F_{aca_1a_2}F_{bda_1a_2}+
\\&+\frac{1}{2}(q-1)\delta_{a,c}\delta_{b,d}\left(\sum_{\eta\neq\alpha}(\lambda^{\beta\eta})^2+\sum_{\eta\neq\beta}(\lambda^{\alpha\eta})^2\right)+\\&
q^{-1}\sum_{\supsubdue{\eta\neq\alpha}{\eta\neq\beta}}\lambda^{\alpha\eta}\lambda^{\eta\beta}\sum_{a_1}v_{aca_1}v_{bda_1}\Bigg)^2 \ ,
\end{split}
\end{equation}
which gives 
\begin{eqnarray}
&&{\bf S}_4^b=\frac{1}{2}(q-1)(q-2)^2\sum_{\alpha\neq\beta}(\lambda^{\alpha\beta})^2+\\
&&\frac{1}{2}(q-1)(q-2)^3\sum_{\alpha\neq\beta}(\lambda^{\alpha\beta})^3+(q-1)(q-2)^2\tr(\lambda^3) \ .\nonumber
\end{eqnarray}
Finally in Eq. \eqref{T4}, we only need the nonzero terms of order $O(1)$ and $O(\lambda)$, \textit{i.e.}, 
\begin{eqnarray}
&&{\bf T}_4=\sum_{\supsubdue{a,b}{c,d}}\delta_{ac}\delta_{bd}\Biggl(\sum_{\supsubtre{\alpha\neq\beta}{\alpha\neq\delta}{\beta\neq\delta}}
Q^{\alpha\beta}Q^{\alpha\delta}\lambda^{\beta\delta}
\\
&&+\frac{1}{2}\sum_{\alpha\neq\beta}(Q^{\alpha\beta})^2
+q^{-2}\frac{1}{2}\sum_{\alpha\neq\beta}(Q^{\alpha\beta})^2\lambda^{\alpha\beta}\sum_{a_1}v_{aca_1}v_{bda_1}\Biggr)\ , \nonumber
\end{eqnarray}
which can be reexpressed as
\begin{equation}
{\bf T}_4=(q-1)^2\sum_{\supsubtre{\alpha\neq\beta}{\alpha\neq\delta}{\beta\neq\delta}}Q^{\alpha\beta}Q^{\alpha\delta}\lambda^{\beta\delta}
+\frac{1}{2}(q-1)^2\sum_{\alpha\neq\beta}(Q^{\alpha\beta})^2 \ .\nonumber
\end{equation}
After putting all these results together in Eq. \eqref{bA2} we find that $-\frac{1}{2}\beta A_2$ is equal to
\begin{equation}
\begin{split}
&\frac{\beta^4}{8d}N(q-1)\Biggl[\left(-2(q-1)+\frac{1}{2}(q-2)^2\right)
\sum_{\alpha\neq\beta}(\lambda^{\alpha\beta})^2 
+\\&
\left(-2(q-1)^2+\frac{1}{2}(q-2)^3\right)
\sum_{\alpha\neq\beta}(\lambda^{\alpha\beta})^3+\\&
+\left(-4(q-1)+(q-2)^2\right)\tr(\lambda^3)
-2(q-1)\tr(Q^2\lambda) \\&
-(q-1)\sum_{\alpha\neq\beta}(Q^{\alpha\beta})^2\Biggr]+
 \\&
 +\frac{\beta^3}{2}\tilde{J}_0N(q-1)(q-2)\bigg[\sum_{\alpha\neq\beta}(\lambda^{\alpha\beta})^2
+(q-2)\sum_{\alpha\neq\beta}(\lambda^{\alpha\beta})^3
+\\&
2\,\tr(\lambda^3)\bigg]+\frac{1}{2}\left(\beta\tilde{J}_0\right)^2dN 
(q-1)\sum_{\alpha\neq\beta}(Q^{\alpha\beta})^2 \ .
\end{split}
\end{equation}

The final piece of information that is needed is the relationship between 
 $\lambda^{\alpha\beta}$ and $Q^{\alpha\beta}$. This results from the equation
 \begin{equation}
\frac{\partial A_0}{\partial \lambda^{\alpha\beta}}=0 \ ,
\end{equation}
which gives
\begin{equation}
Q^{\alpha\beta}=\lambda^{\alpha\beta}+\frac{1}{2}(q-2)(\lambda^{\alpha\beta})^2+\sum_{\eta}\lambda^{\alpha\eta}\lambda^{\eta\beta} \ ,
\end{equation}
\begin{equation}
\lambda^{\alpha\beta}=Q^{\alpha\beta}-\frac{1}{2}(q-2)(Q^{\alpha\beta})^2-\sum_{\eta}Q^{\alpha\eta}Q^{\eta\beta}
\end{equation}
and
\begin{equation}
(\lambda^{\alpha\beta})^2=(Q^{\alpha\beta})^2-(q-2)(Q^{\alpha\beta})^3-2Q^{\alpha\beta}\sum_{\eta}Q^{\alpha\eta}Q^{\eta\beta} \ .
\end{equation}
For the second-order term we therefore have
\begin{eqnarray}
&&-\frac{1}{2}\beta A_2=N(q-1)\Biggl[\Biggr.
\sum_{\alpha\neq\beta}(Q^{\alpha\beta})^2\left(
\frac{\beta^4}{16d}(q^2-10q+10)+\right.\Biggr.\nonumber\\
&&\left.\Biggl.\frac{\beta^3}{2}\tilde{J}_0(q-2)+\frac{1}{2}(\beta\tilde{J}_0)^2d\right)+\Biggr.\\\Biggl.
&&\nonumber
-\sum_{\alpha\neq\beta}(Q^{\alpha\beta})^3\frac{\beta^4}{4d}(q-1)
-\tr(Q^3)\frac{\beta^4}{4d}(q-1)
\Biggl.\Biggr] \ ,
\end{eqnarray}
while for the zeroth-order term we obtain 
\begin{equation}
\begin{split}
&-\beta A_0=N(p-1)\bigg[-\frac{1}{4}\sum_{\alpha\neq\beta}(Q^{\alpha\beta})^2 \\&
+\frac{1}{12}(q-2)\sum_{\alpha\neq\beta}(Q^{\alpha\beta})^3 +\frac{1}{6}\tr(Q^3)\bigg] \ ,
\end{split}
\end{equation}
and for the first-order term,
\begin{equation}
-\beta A_1=N(q-1)\frac{\beta^2}{4}\sum_{\alpha\neq\beta}(Q^{\alpha\beta})^2 \ .
\end{equation}
After collecting all the pieces together, we finally find (formally setting $\upsilon=1$) 
\begin{eqnarray}
&&-\beta A\simeq N(q-1)\Biggl[
\sum_{\alpha\neq\beta}(Q^{\alpha\beta})^2\left(\frac{\beta^2}{4}-
\frac{1}{4}+\right. \Biggr.\\
&&\left.\Biggl.\frac{\beta^4}{16d}(q^2-10q+10)+
\frac{\beta^3}{2}\tilde{J}_0(q-2)+
\frac{1}{2}(\beta\tilde{J}_0)^2d\Biggr.\right)+\nonumber\\&&\Biggl.
\nonumber
\Biggr.+\sum_{\alpha\neq\beta}(Q^{\alpha\beta})^3\left(
\frac{1}{12}(q-2)-\frac{\beta^4}{4d}(q-1)
\right)+\Biggr.\\\Biggl.
&&
\tr(Q^3)\left(
\frac{1}{6}-\frac{\beta^4}{4d}(q-1)
\right)
\Biggr] \ .\nonumber
\end{eqnarray}

\section{Mean-field equations for the $M$-$p$ spins model}

As in the main text, we focus on the $M=3$ and $p=3$ model.  In this case, by following the standard procedure described in Ref. [\onlinecite{spinglassbeyond}],
one obtains a mean-field solution in terms of the two overlaps $Q^{(1)} $ and $Q^{(2)}$
introduced in the text.  Below, we only reproduce the final result. 
Let us first introduce the notation
\begin{eqnarray}
&&{\Huge \nu}(\beta,Q^{(1)},Q^{(2)},\{S^{\alpha}\})=\\
&&\exp\left[\sum_{\alpha}\beta\sqrt{Q^{(2)}}z^{\alpha}_2S^{\alpha}+\sum_{\alpha}\beta\sqrt{\displaystyle\frac{Q^{(1)}}{3}}z_1^{\alpha}
\prod_{\gamma\neq\alpha}S^{\gamma}\right] \ . \nonumber
\end{eqnarray}
The self-consistent equations for the two overlaps read:
\begin{equation}
\begin{aligned}
&Q^{(1)}=\\&
\frac{\displaystyle\int\prod_{\alpha=1}^3\frac{dz_1^{\alpha}}{\sqrt{2\pi}}\frac{dz_2^{\alpha}}{\sqrt{2\pi}}\displaystyle\frac{
    \left\{\sum_{\{S\}}\Huge{\nu}(\beta,Q^{(1)},Q^{(2)},\{S^{\alpha}\})S^1\right\}^2}{                                                                                                                                                   
    \sum_{\{S\}}\Huge{\nu}(\beta,Q^{(1)},Q^{(2)},\{S^{\alpha}\})}}{8\exp\left[\beta^2\left(\displaystyle\frac{Q^{(1)}+3Q^{(2)}}{2}\right)\right]}
\end{aligned}
\end{equation}

\begin{equation}
\begin{aligned}
&Q^{(2)}=\\&
\frac{\displaystyle\int\prod_{\alpha=1}^3\frac{dz_1^{\alpha}}{\sqrt{2\pi}}\frac{dz_2^{\alpha}}{\sqrt{2\pi}}\displaystyle
\frac{\left\{\sum_{\{S\}}\Huge{\nu}(\beta,Q^{(1)},Q^{(2)},\{S^{\alpha}\})S^1S^2\right\}^2}{                                                                                                                                                   
    \sum_{\{S\}}\Huge{\nu}(\beta,Q^{(1)},Q^{(2)},\{S^{\alpha}\})}}{8\exp\left[\beta^2\left(\displaystyle\frac{Q^{(1)}+3Q^{(2)}}{2}\right)\right]} \ .
\end{aligned}
\end{equation}

The free energy [divided by $(m-1)$] is given by
\begin{equation}
\begin{aligned}
&F=\frac{3}{2}\beta^2Q^{(1)}Q^{(2)}+\frac{3}{2}\beta^2Q^{(2)}+\frac{1}{2}\beta^2Q^{(1)}+\log(8)
\\&-\displaystyle\frac{\displaystyle\int\prod_{\alpha=1}^3\frac{dz_1^{\alpha}}{\sqrt{2\pi}}\frac{dz_2^{\alpha}}{\sqrt{2\pi}}
  \Huge{\nu} \log \Huge{\nu}(\beta,Q^{(1)},Q^{(2)},\{S^{\alpha}\})}{8\exp\left[\beta^2\left(\displaystyle\frac{Q^{(1)}+3Q^{(2)}}{2}\right)\right]} \,\, .
\end{aligned}
\end{equation}

\section{Mean-field equations for the F-model}

We assume a 1-RSB ansatz for the matrix $Q^{\alpha \beta}$ and let $Q$ denote the intra-state overlap.
The corresponding mean-field equations can be derived by usual methods and read
\begin{equation}
Q=\frac{\int_{-\infty}^{+\infty}\prod_{a=1}^3dz_a\frac{e^{-z_a^2/2}}{\sqrt{2\pi}}\frac{
\left[\sum_{\mathcal{S}}\mu'(\mathcal{S})S^1\exp\left(\beta \sqrt{Q}\sum_a S_a z_a\right)
 \right]^2
}{\sum_{\mathcal{S}}\mu'(\mathcal{S})\exp\left(\beta \sqrt{Q}\sum_a S_a z_a\right)
}}
{\sum_{\mathcal{S}}\mu'(\mathcal{S})\exp\left(\beta^2 \frac Q 2 \sum_a (S_a)^2\right)}
\end{equation}
where $\mathcal{S}$ is a short-hand notation for $S_1,S_2,S_3$ and 
$\mu'(\mathcal{S})=\mu(\mathcal{S})\exp\left[\beta^2\left(u-\frac{Q}{2} \right)\sum_a(S_a)^2 \right]$.
The effective field $u$ has to be determined self-consistently from the equation 
\begin{equation}
\begin{aligned}
u&=\frac 1 2 \int_{-\infty}^{+\infty}\prod_{a=1}^3dz_a\frac{e^{-z_a^2/2}}{\sqrt{2\pi}}\times\\
&\frac{\sum_{\mathcal{S}}\mu'(\mathcal{S})(S_1)^2\exp\left(\beta \sqrt{Q}\sum_a S_a z_a\right)
}{\sum_{\mathcal{S}}\mu'(\mathcal{S})\exp\left(\beta^2 \frac Q 2 \sum_a (S_a)^2\right)} \\&
= \frac 1 2\, \frac{\sum_{\mathcal{S}}\mu(\mathcal{S})(S_1)^2\exp\left(\beta^2 u \sum_a (S_a)^2\right)
}{\sum_{\mathcal{S}}\mu(\mathcal{S})\exp\left(\beta^2 u \sum_a (S_a)^2\right)} \, .
\end{aligned}
\end{equation}
By solving numerically these equations we have obtained the results described in the main text.
In the high-temperature, paramagnetic, region regime where $Q=0$, on-site spin averages can be 
obtained by just using the measure $\mu'(\mathcal{S})$. In this way, one can compute the correlation functions
entering in the expansion of the Gibbs free energy.


\begin{thebibliography}{99}
\bibitem{footnote1}
Since $\pi_{ij}^{-1}=\pi_{ij}$ the interaction $\delta(\sigma_i, \pi_{ij}(\sigma_j))$ is equivalent to 
$\delta(\pi_{ij}(\sigma_i), \sigma_j)$. Therefore the interaction is symmetric with respect to the exchange 
of the sites $i$ and $j$.
\bibitem{footnote2}
Note that the threshold $c_{col}$ for the existence of a coloring transition at $T=0$ is slightly 
larger than the limiting RFOT $c_K$ as it is such that\cite{krzakala07} $c_{col}\sim 2q\log(q) - \log(q)-1$.
\bibitem{footnote3}
On a regular random graph, long-range antiferromagnetic order cannot truly exist. However, one can still calculate the stability of the 
paramagnetic solution and assign a direction of instability to incipient antiferromagnetism.
\bibitem{footnote3b}Within replica theory, assuming that the Almeida-Thouless transition line exists, the 
amplitude of this tail would go to zero at the transition. Within droplet theory, long-range order is destabilized by the field on a length-scale $\xi_h$
proportional to $h^{-1/(d/2-\theta)}$, see Ref. \onlinecite{droplet_fisher}. The overlap would first decrease toward a value $q_1$ on length-scales of the order of one and then 
to a value $q_0$ for cavity sizes larger than $\xi_h$. However, both close to the remnant of the Almeida-Thouless line and close to $h=0$ the difference
$q_1-q_0$ is expected to be small. In consequence, the growth of the point-to-set length would only be visible in the tail behavior.
\bibitem{footnote4}
However, one should not conclude that studying disordered fully connected models is always worthless as far 
as the glass transition is concerned. They may indeed fare better for describing glass-forming liquids than their own finite-dimensional 
version. This---apparently odd---statement comes from the fact discussed in Sec. VI that models of glass-forming liquids may, 
after some coarse-graining, behave more as predicted from infinite-dimensional RFOT models than finite-dimensional 
disordered spin models do. The main advantage then of infinite-dimensional/fully-connected models is that they can be rigorously 
analyzed in depth and that new effects and phenomena can be more easily looked for.
\end{thebibliography}
\end{document}